\documentclass[twocolumn,showpacs,english,superscriptaddress,preprintnumbers,amsmath,amssymb,floatfix]{revtex4-1}

\usepackage[T1]{fontenc}
\usepackage[latin9]{inputenc}
\usepackage{dcolumn}
\usepackage{bm}
\usepackage{graphicx}
\usepackage{color}
\usepackage{esint}
\usepackage{babel}
\usepackage{amsfonts}
\usepackage{slashed}
\usepackage{enumerate}

\begin{document}

\title{Roadmap to Majorana surface codes}

\author{S.~Plugge}
\affiliation{Institut f\"ur Theoretische Physik,
Heinrich-Heine-Universit\"at, D-40225  D\"usseldorf, Germany}
\affiliation{Center for Quantum Devices and Station Q Copenhagen,  
Niels Bohr Institute, University of Copenhagen, 
Universitetsparken 5, DK-2100 Copenhagen, Denmark}

\author{L.A.~Landau}
\affiliation{Raymond and Beverly Sackler School of Physics and Astronomy,
Tel-Aviv University, Tel Aviv 69978, Israel}

\author{E.~Sela}
\affiliation{Raymond and Beverly Sackler School of Physics and Astronomy,
Tel-Aviv University, Tel Aviv 69978, Israel}

\author{A.~Altland}
\affiliation{Institut f\"ur Theoretische Physik,
Universit\"at zu K\"oln, Z\"ulpicher Str.~77, D-50937 K\"oln, Germany}

\author{K.~Flensberg}
\affiliation{Center for Quantum Devices and Station Q Copenhagen,  
Niels Bohr Institute, University of Copenhagen, 
Universitetsparken 5, DK-2100 Copenhagen, Denmark}

\author{R.~Egger}
\affiliation{Institut f\"ur Theoretische Physik,
Heinrich-Heine-Universit\"at, D-40225  D\"usseldorf, Germany}
\date{\today}

\begin{abstract}
Surface codes offer a very promising avenue towards fault-tolerant quantum 
computation. We argue that two-dimensional interacting networks of Majorana bound states in topological superconductor/semiconductor heterostructures hold several distinct advantages in that direction, both concerning the hardware realization and the actual operation of the code.   We here discuss how topologically protected logical qubits in this Majorana surface code architecture can be defined, initialized, manipulated, and  read out.  All physical ingredients needed to implement these operations are routinely used in topologically trivial quantum devices.  In particular, we show that by means of  quantum interference terms in linear conductance measurements, composite single-electron pumping protocols, and  gate-tunable tunnel barriers, 
the full set of quantum gates required for universal quantum computation can be implemented. 
\end{abstract}

\pacs{03.67.Pp, 03.67.Ac, 03.67.Lx, 74.78.Na}

\maketitle

\section{Introduction}\label{sec1}

Surface code architectures are currently being recognized as potentially very
powerful platforms for quantum information processing (QIP) and universal quantum
computation
\cite{Gottesman1997,Bravyi2001,Freedman2001,Dennis2002,Bravyi2005,Raussendorf2007,
Raussendorf2007b,Fowler2011,Fowler2012a,Fowler2012,Terhal2015}.  Based 
on relatively simple two-dimensional (2D) array structures, the distinguishing 
hallmark of code-based QIP is redundancy: a rather small number of logical (or information) qubits
 is integrated into the background of a much larger number of physical qubits.   
The latter are built from local combinations of elementary hardware qubits and act as so-called stabilizers of the code, 
i.e., they are subject to readout operations projecting the system to a well-defined code state.
The encoding of relatively few logical qubits into a large entangled code space provides a highly potent resource for error
detection/correction and fault tolerance in quantum operations.  In fact, as long as the microscopic error probability 
for physical qubit manipulations stays below a rather benign threshold of order $\approx 1 \%$ \cite{Fowler2011,Fowler2012a,Fowler2012,Terhal2015}, 
logical qubits enjoy a topological protection with exponentially small error rates for increasing ratio of physical to logical qubits \cite{Dennis2002}.  
Somewhat less obvious, this encoding strategy also makes the boundaries between hard- and software less rigid than in other QIP approaches. 
For example, whether a given physical qubit acts as stabilizer or as storage space for logical information in each computation cycle will be 
decided by the code protocols and the applied measurement pattern.  To a certain degree, error handling can then be reduced to simply
tracking errors by classical software.  An excellent and detailed review of the surface code approach to quantum computing
has been given by Fowler \textit{et al.} \cite{Fowler2012}. 
 
However, there is a price to be paid for the high levels of flexibility and fault
tolerance in QIP based on surface codes, namely the tremendous hardware overhead
introduced by the code space.  Referring to Refs.~\cite{Fowler2011,Fowler2012a,Fowler2012} for detailed estimates, 
the realization of a useful system (allowing for well-protected QIP) of just a few dozen logical qubits might require
to assemble thousands of physical qubits.
This means that maximal efficiency will be critically important to the design of scalable surface code platforms, 
raising questions both concerning the physical realization of the code and, equally important, its operation:

\noindent \emph{Realization.---} What levels of physical qubit integration can be
reached at  nanoscopic or mesoscopic length scales? How much hardware overhead is
required to define, manipulate, and read out logical qubit states? To what extent is
the manipulation of individual qubits detrimental to the operation of the code, say, with 
regard to generated heat or radiation levels?   If external
electromagnetic fields are necessary, can their cumulative effects be kept at bay
when large numbers of qubits are integrated?

\noindent \emph{Operation.---} Can one address qubits with logarithmic efficiency \cite{Terhal2015}, e.g., via non-invasive multiplexer schemes? 
 Recalling that code-based QIP relies on the repeated readout of many stabilizers,
 can such measurements be performed at sufficient speed, say, with a cycle time of a
 few microseconds? Can the full set of logical gates required for universal quantum
 computation be realized at tolerable efforts? How can one avoid decay and dephasing
 of qubits due to residual interactions, e.g., with readout or manipulation devices,
 or between qubits themselves?

In Ref.~\cite{Aviad2016}, we have argued that topological hybrid
superconductor/semiconductor architectures featuring Majorana bound states might
evolve into a platform performing favorably with regard to many of the above
criteria.  (For reviews on Majorana states in such devices, see
Refs.~\cite{Alicea2012,Leijnse2012,Beenakker2013,DasSarma2015}. Majorana surface
codes are also discussed in
Refs.~\cite{Terhal2010,Terhal2012,VijayFu2015,VijayFu2016}.) Given that not even a
single topological semiconductor qubit has been realized so far, this assertion may
seem presumptuous. However, current striking developments in the materials
foundations of topological semiconductor devices, including experimentally observed
signatures of Majorana states, see Refs.~\cite{Kouwenhoven2012,Chang2015,
Krogstrup2015,Higginbotham2015,Albrecht2016,Leo2016,Marcus2016}, give rise to cautious
optimism that the realization of functioning device architectures is but a matter of
time. Two developments in particular seem of key importance. First, progress in the
growth of high-quality superconductor/semiconductor interfaces has been instrumental
to the formation of hard excitation gaps and to the protection of Majorana states
\cite{Chang2015,Krogstrup2015,Higginbotham2015,Albrecht2016,Leo2016}. Second, the
successful realization of 2D superconductor/semiconductor heterostructures
\cite{Marcus2016,Kjaergaard2016} suggests that it may be advantageous to switch from nanowire
architectures to genuine 2D layouts. For concreteness, we describe the wire 
construction below. However, we do not expect that the analogous formulation of a 
2D architecture will differ on a conceptual level. 
Once realized, either structure will be accessible in terms of the full spectrum of device manipulations and measurement
protocols of mesoscopic quantum electronics, including (i) quantum interferometry by
point contact measurements of the linear conductance,  (ii) access by quantum dots
that allow one to pump single electrons on and off
the code, and (iii) the option to connect and disconnect qubits from the code by
tuning selected voltages.

\begin{figure*}
\centering
\includegraphics[width=12.5cm]{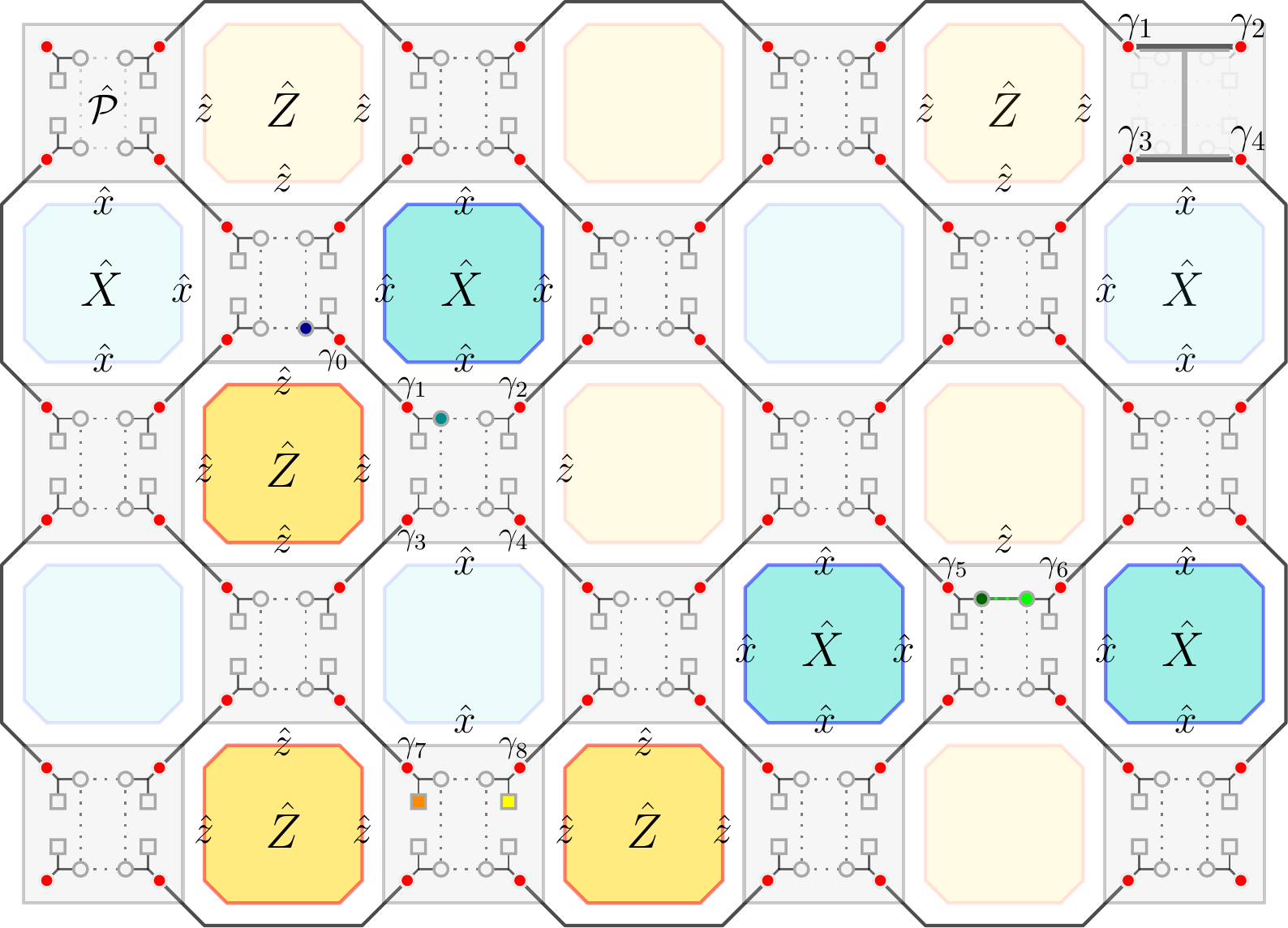}
\caption{Schematic Majorana code architecture with 18 MCBs (light grey). 
The MCB is a floating superconducting island with four Majorana end states $\gamma_{j=1,\ldots,4}$ (red dots) 
of topological semiconductor wires, which can be implemented, e.g., as 90-deg rotated H-shaped device 
\cite{Aviad2016} (for an illustration, see top right corner).
 The fermion parity on a given MCB,
 $\hat{\cal P}=\gamma_1\gamma_2\gamma_3\gamma_4$, is constrained by Coulomb 
blockade, see Eq.~\eqref{parityconstraint}, and 
Majorana bilinears define Pauli operators $(\hat x,\hat z)$, see Eq.~\eqref{PauliMajorana}.    
Majorana states on neighboring MCBs are linked by tunnel bridges (solid black lines).  
Away from (near) boundaries, minimal plaquettes involve products of eight (six) Majoranas. 
These stabilizers are either of type $\hat X$ (blue) or $\hat Z$ (yellow), corresponding to 
products of four (three) $\hat x$ or $\hat z$ operators, resp., 
see Eq.~\eqref{stabilizerOp2}.  Note that near the top/bottom $Z$-type (left/right $X$-type)
 code boundaries, only $\hat Z$ ($\hat X$) stabilizers appear.  
Stabilizer eigenvalues are read out by conductance measurements, see Eq.~\eqref{conductance}.  
By using leads (small circles) tunnel-coupled to 
$\gamma_0$ and $\gamma_1$, resp., one measures the two adjacent (colored) stabilizers. Majorana 
bilinears can be read out by adding interference links 
between leads (dotted lines).  By using leads attached to $\gamma_5$ and $\gamma_6$, we can measure 
$\hat z = i\gamma_5\gamma_6$, cf.~Eq.~\eqref{conductance2}. 
This Pauli operator readout affects the neigboring (blue) $\hat X$ stabilizers, which anticommute with $\hat z$.
 By pumping a single electron between two quantum dots (indicated by 
squares), one generates a string operator.  With dots attached to $\gamma_7$ and $\gamma_8$,
 the latter is $\hat x = i\gamma_7\gamma_8$ and flips 
the two adjacent (yellow) $\hat Z$ stabilizers. }
\label{f1}
\end{figure*}

To make these statements more concrete, let us start with a brief synopsis of the key features
of a  Majorana surface code, see also Sec.~\ref{sec2} below. Its
skeleton structure is defined by a regular square lattice of so-called
Majorana-Cooper boxes (MCBs) linked by tunnel bridges, see Fig.~\ref{f1}. 
Each MCB corresponds to two topological semiconductor nanowires proximitized by a 
common, floating superconducting island. The nanowires in the setup of
Fig.~\ref{f1} are parallel, which means  that a homogeneous parallel magnetic Zeeman
field  can synchronously drive them into a topological phase  
\cite{Alicea2012,Leijnse2012,Beenakker2013}. Within the topological phase, each wire supports Majorana
zero-energy end states, giving rise to a total of four states per MCB. 
Each Majorana state is assumed to be individually tunnel-coupled to an external lead electrode, and to a
quantum dot (or single-electron transistor). We note that the implementation of these
access units does not require significant hardware beyond that defining the MCBs as
such.  Below we will augment  the hardware described in Ref.~\cite{Aviad2016} by
local interference links between nearby pairs of external leads. Such links can be realized as
gate-tunable tunnel bridges that may be switched on and off as needed. Once in place, 
they will allow for interferometric transport measurements of 
Majorana strings (e.g., bilinears), which will in turn be instrumental to the manipulation 
of logical operators. We note that topological T-junctions  required, e.g., for the braiding of Majorana
fermions \cite{Aasen2016} are not needed in our approach.

Assuming a large capacitive charging energy, backgate voltages may be tuned to effect
charge quantization on each MCB \cite{Fu2010,Zazunov2011,Plugge2015}.  
The low-energy state of a given MCB is then equivalently described by an 
effective spin-1/2 degree of freedom
\cite{BeriCooper2012,AltlandEgger2013,Beri2013,Zazunov2014,Tsvelik2014,Plugge2016}.
These spin-$1/2$ variables are the elementary hardware qubits of our system
and can be described by Pauli matrices $(\hat x,\hat y,\hat z)_l$ for MCB no.~$l$.
 We assume tunnel bridges linking neighboring MCBs  into the checkerboard pattern of
Fig.~\ref{f1}. This coupling generates spin-$1/2$ ring-exchange processes around the
elementary plaquettes (loops) of the 2D lattice which may be described as products
of MCB Pauli operators, either of four $\hat z_l$ or of four $\hat x_l$.  The emerging two types
of composite physical qubits, denoted by $\hat Z$ and $\hat X$ in Fig.~\ref{f1}, are
the stabilizers of the surface code. Physically, one may think of the  stabilizer qubit  as the
 $\Bbb{Z}_2$-valued flux captured by the lattice ring-exchange processes. 

Our basic approach to the repeated stabilizer measurement is quantum interferometry
 (cf.~Refs.~\cite{Bonderson2007,Bonderson2008}):
 Current passing through two terminals adjacent to a plaquette can pass around this loop in clockwise 
 or anti-clockwise direction, and the interference between these paths identifies qubit states through conductance 
 measurements. This single-step measurement protocol neither requires ancilla qubits, 
 cf.~also Refs.~\cite{VijayFu2015,VijayFu2016}, nor hardware beyond that already present in the system.  
 Via the above-mentioned interference links, similar interferometric transport measurements can determine the 
 eigenvalues of arbitrary Majorana bilinears and strings, including those of the 
qubit operators $(\hat x,\hat y,\hat z)_l$. 

We have proposed the basic access hardware and sketched how to perform  
elementary code operations in Ref.~\cite{Aviad2016}, including stabilizer readout and the 
controlled creation of anyonic excitations.  Here we take the crucial next step 
and demonstrate how the semiconductor-based code architecture may become a 
potent platform for QIP. This includes  the initialization, manipulation, protection, and 
readout of logical qubits in most general
terms. In doing so, we try to strike a middle ground, avoiding excessive levels of
material-specific details, yet aiming to be specific enough to let the reader judge on
the concrete perspectives of the approach. Our goal is to provide a comprehensive
introduction to QIP in this code architecture, up to and including the
implementation of a minimal set of gates  for universal quantum computation.

The information content held by a surface code depends on its effective
ground-state degeneracy, which in turn is governed by the number of holes in the
code. Holes can be effectively created by excluding individual stabilizers from the
readout operation applied to the majority of physical qubits in each computation cycle.   
This solution, which resembles the software hole of Ref.~\cite{Fowler2012}, may be implemented 
by only switching on (via gate voltage changes) the respective tunnel couplings 
between probe leads and the code when a readout is performed. 
We note that it is not sufficient to simply turn off the respective bias voltage in the idle state.
For instance, unbiased leads can still decohere the system. 
Hardware holes may be created by gate-blocking  the intra-code connections along a loop which
defines a chosen stabilizer, or a group of stabilizers. In general, it will be
necessary to employ hybrid strategies comprising elements of both the soft- and
hardware realization of holes. For example, when a stabilizer is inactive or remains
idle, it may be advantageous to ramp down its connection to the rest of the code so as
to minimize unwanted perturbations. The coupling can then be turned on again prior to
initialization, readout, manipulation, or related operations.

It is obvious that experimental efforts will first focus on structures simpler than
that of a large 2D code.  This includes 1D chains that may implement so-called  
repetition codes \cite{Martinis2015} which do not achieve the full computational power of a 2D surface code. 
Nonetheless, many of the manipulation and readout tools needed for QIP 
in the 2D code can already be implemented and tested using small MCB chains or arrays.
In a related work \cite{paperB}, we shall discuss minimal models composed of just a few
tunnel-coupled MCBs. We also note that the Majorana surface code approach 
does not rely on the non-Abelian braid operations 
underlying alternative approaches to Majorana-based quantum computing
\cite{Sau2010,Alicea2011,Clarke2011,Flensberg2011,
Leijnse2011,Bonderson2011,Hyart2013,Pedrocchi2015,Clarke2016,Loss2016},
cf.~Ref.~\cite{VijayFu2015} for a detailed discussion. 
All its operations are performed through conductance 
measurements, pumping protocols, and/or gate voltage changes.  
Our logical gate implementations are  different, and sometimes much simpler, than those suggested for bosonic codes 
\cite{Fowler2012,Barends2014,Jeffrey2014,Corcoles2015,Martinis2015,Diamond2015}, 
other Majorana code realizations \cite{Terhal2012,VijayFu2015,VijayFu2016}, 
or platforms relying on non-Abelian braiding.  
Given the challenging demands imposed by scalability \cite{Terhal2015}, we 
believe that these signatures may be of critical importance for the 
successful realization of large-scale systems.
  
The remainder of this paper is structured as follows. We start out in Sec.~\ref{sec2}
with a discussion of the different hardware hierarchies defining our code
architecture. In order to keep the paper self-contained, Sec.~\ref{sec2} also
summarizes the key results of Ref.~\cite{Aviad2016}. In Sec.~\ref{sec3}, we
show how logical qubits are defined.  Basal quantum operations
are discussed in  Sec.~\ref{sec4}, where we 
demonstrate how logical qubits can be initialized, moved, braided, manipulated, 
and read out in this architecture. To that end, we will also introduce a powerful class
 of generalized pumping protocols.  In Sec.~\ref{sec4f}, we offer an analysis of potential
challenges and/or error sources.  With those basal operations at hand, the 
implementation of a full set of quantum gates for universal quantum computation 
is possible using established quantum circuits 
\cite{Fowler2012,VijayFu2015,Nielsen2000}, see Sec.~\ref{sec5}.  
The paper ends with concluding remarks in Sec.~\ref{sec6}.  
Finally, we note that we often use units with $\hbar=1$ below.

\section{Majorana surface code hardware} \label{sec2}

\subsection{Majorana-Cooper box}\label{sec2a}

The elementary unit of our architecture is the Majorana-Cooper box (MCB), described in detail in Refs.~\cite{BeriCooper2012,AltlandEgger2013,Beri2013,Zazunov2014,Tsvelik2014,Plugge2016}
and schematically illustrated in Fig.~\ref{f1}. In the following we briefly introduce the MCB, 
where the emphasis will be on its subsequent integration into a surface code 
structure. Interested readers may find a concrete proposal for the hardware layout of an 
 MCB in Ref.~\cite{Aviad2016}. 
An isolated MCB consists of a mesoscopic $s$-wave superconducting island in contact
to two semiconductor nanowires with strong spin-orbit interaction  (e.g., InAs or
InSb). On each wire a pairing gap is induced by the proximity effect, 
which we assume to be so large that continuum quasiparticles can be ignored.
Residual quasiparticle poisoning effects are discussed in Sec.~\ref{sec4f}. 
When subjected to a sufficiently strong Zeeman field, the wires  enter a topological regime with
Majorana bound states  near their ends \cite{Alicea2012,Leijnse2012,Beenakker2013}.
We here  consider  wires long enough  that direct tunnel couplings between different end
states are negligible, but finite-coupling effects are briefly addressed in Sec.~\ref{sec4f}. 

In effect, as illustrated in Fig.~\ref{f1}, we then have four zero-energy Majorana states described by
Hermitian operators, $\gamma^{}_j=\gamma_j^\dagger$, which satisfy  a Clifford
anticommutator algebra, $\{
\gamma_j,\gamma_{j'}\}=2\delta_{jj'}$ \cite{Alicea2012,Leijnse2012,Beenakker2013}.  
Recalling that a conventional fermion can be built from two Majorana fermions, the presence of
four zero-energy Majorana states  implies a four-fold degeneracy.  This degeneracy
gets lifted by the  Coulomb blockade, which has to be taken into account because we
consider a floating (not grounded) island with a large single-electron charging
energy  $E_C\approx 1$~meV \cite{Albrecht2016}. For notational simplicity, we shall 
consider the same $E_C$ for all MCBs below.   In what follows, the effective energy scales
of the problem are assumed to be below both the proximity-induced pairing gap and
the charging energy.

Since Cooper pairs and Majorana fermions both represent zero-energy states, the 
Hamiltonian of a single MCB contains only the capacitive charging energy contribution,
$H_{\rm MCB}= E_C(\hat N-n_g)^2,$ where the parameter $n_g$ can be tuned by a backgate voltage. 
 The operator $\hat N$ has integer eigenvalues ($N$) and counts the fermion number on the MCB.  
 Within our low-energy framework, these fermions are either bound in Cooper pairs or belong to the Majorana sector.    
 Under Coulomb valley conditions, i.e., for $n_g$ close to integer values, the charging energy 
 then enforces a well-defined ground-state value for $N$, even if weak tunnel couplings to other 
 code parts are present. Depending on our choice for $n_g$, this value is either even or odd. 
 Since the number of electrons bound in  Cooper pairs is always even, 
 a parity constraint emerges in the Majorana sector \cite{BeriCooper2012}, 
\begin{equation}\label{parityconstraint}
\hat{\cal  P} = \gamma_1\gamma_2\gamma_3\gamma_4=(-1)^N=\pm 1,
\end{equation} 
reducing the four-fold to a two-fold degeneracy. With charge fluctuations being frozen out by Coulomb blockade,
the low-energy dynamics of the MCB is then described by an effective spin-$1/2$ variable.
The associated Pauli operators $(\hat x,\hat y,\hat z)$ correspond to bilinear combinations of Majorana operators  
and afford the representation \cite{BeriCooper2012,Tsvelik2014,Plugge2016}  
\begin{equation}\label{PauliMajorana}
\hat x = i \gamma_1\gamma_2, \quad \hat y =i \gamma_2 \gamma_3, \quad
\hat z = i \gamma_1 \gamma_3.
\end{equation}
Note that this representation is consistent with standard Pauli matrix relations, $\hat x \hat y=i\hat z$ and so on.
Using the parity constraint \eqref{parityconstraint}, each Pauli operator can equivalently be expressed by the complementary pair of Majorana operators, e.g., 
$\hat x = \mp i\gamma_3\gamma_4$.   For 2D arrays of MCBs, it is convenient to choose an alternating representation as illustrated in Fig.~\ref{f1}.  For MCB no.~$l$, 
Pauli operators ($\hat x_l,\hat y_l,\hat z_l)$ are assigned for $l$ in an odd column (containing three MCBs in Fig.~\ref{f1}),
but the unitarily transformed representation $(\hat z_l,-\hat y_l, \hat x_l)$ will be used for even columns (with two MCBs in Fig.~\ref{f1}).
This convention leads to a maximally transparent representation of stabilizer operators in terms of MCB 
Pauli operators, see Eq.~\eqref{stabilizerOp2} below. 

The tunnel coupling between Majorana state no.~$j$ of an MCB and an attached lead or quantum dot with fermion operator $\Psi$
is described by the tunneling Hamiltonian \cite{Fu2010,Zazunov2011,Plugge2016}
\begin{equation}\label{TunnelingOperator}
H_{\rm t} = t_j \Psi^\dagger e^{-i\hat \varphi/2} \gamma_j +\text{h.c.},
\end{equation}
where $t_j$ is a complex tunneling amplitude and the phase operator $\hat \varphi$ is 
conjugate to the number operator, $[\hat \varphi,\hat N]=2i$. The factor of two means that
$e^{i\hat \varphi}$ creates a Cooper pair, $N\to N+2$, and hence the 
out-tunneling process contained in Eq.~\eqref{TunnelingOperator} annihilates a single electron 
charge on the island.   The representation \eqref{TunnelingOperator} manifestly splits the 
tunneling process into a transfer of charge and the charge-neutral Majorana dynamics.

\subsection{MCB network}\label{sec2b}

We next consider a  network of MCBs as shown in Fig.~\ref{f1}, where MCB no.~$l$ has
the fluctuating superconducting phase $\varphi_l$. Neighboring MCBs are  coupled
through short tunnel bridges connecting specific pairs of Majorana states
$(\gamma_l,\gamma_{l'})$. Describing each contact in terms of the tunneling
Hamiltonian \eqref{TunnelingOperator} with tunnel amplitude $t_l$ between $\gamma_l$
and the respective bridge fermion $\Psi_l$, the effective MCB-MCB tunneling
Hamiltonian follows by second-order perturbation theory,
 \begin{equation} \label{TunnelingMCB}  
H_{\rm tun}= -\frac{t_{ll'}}{2}e^{i(\hat\varphi_l-\hat\varphi_{l'})/2} \gamma_{l}\gamma_{l'}+{\rm h.c.},
\end{equation}
where $t_{ll'} \simeq t^{\ast}_l t^{}_{l'} G(l,l')$ contains the low-energy fermion propagator of the bridge. 
 For a sufficiently short bridge, retardation effects in this propagator are negligible and 
 the coupling constants $t_{ll'}$ are effectively time- and/or energy-independent.  
As an example with the above properties, one may think of tunnel bridges that include short 
non-proximitized nanowire segments connecting adjacent MCBs. Such small-sized 
connectors typically form junction dots, i.e., few-level systems where electrostatic gating makes the $t_{ll'}$ couplings tunable.
For $|t_{ll'}|\agt 0.33 E_C$, the ground state of the system enters a frustrated 2D Ising phase which is not useful for QIP purposes \cite{Terhal2012}. 
We therefore focus on the regime  $|t_{ll'}|\ll E_C$ throughout. 

The presence of the exponentiated phase operators in Eq.~\eqref{TunnelingMCB} indicates
 that a single tunneling process kicks both participating MCBs out of their respective charge ground states, 
 leaving the system in an excited charge state with energy of order $E_C$.  
  One option to go back to the charge ground state is given by a trivial reverse tunneling event 
  along the same link which only adds an irrelevant additive constant to the effective low-energy 
  code Hamiltonian.  
Dropping such constants, the leading-order process for relaxing back to the charge ground
 state comes from 
ring exchange along the elementary plaquettes involving four MCBs. 
The integration over the transient times of order $E_C^{-1}$ where the system is virtually excited
is most conveniently done by applying a Schrieffer-Wolff transformation \cite{BravyiLoss2011}. 
 We then obtain the  effective low-energy code Hamiltonian \cite{Aviad2016} 
\begin{equation}\label{Sysham}
H_{\rm code} = - \sum_n  {\rm Re}(c_n) \hat {\cal O}_n,
\quad c_n=  \frac{5}{16 E_C^3} \prod t_{l_n^{} l_n^\prime},
\end{equation}
where $n$ runs over all elementary plaquettes of the 2D network. 
Ignoring the code boundaries for the moment, this defines stabilizer operators $\hat {\cal O}_n$ as products of the eight Majorana operators comprising a minimal loop,
$\hat {\cal O}_n= \prod_{j=1}^8 \gamma_j^{(n)}$.
The coefficients $c_n$ in Eq.~\eqref{Sysham} involve the product of the four tunnel amplitudes around this loop.
Note that the fluctuating phases $\varphi_l$ have dropped out of Eq.~\eqref{Sysham} due to the Schrieffer-Wolff projection to the charge ground state of each MCB.
A key feature of the Hamiltonian~\eqref{Sysham} is that all contributing operators commute, $[ \hat{\cal O}_n,\hat{\cal O}_{n'}]=0$, since two plaquettes
either share two or no Majorana operators.  Each stabilizer operator  
 is Hermitian, $\hat{\cal O}^{}_n=\hat{\cal O}_n^\dagger$, and squares to unity, $\hat {\cal O}_n^2=1$. The Hilbert spaces 
defined by the two possible eigenvalues $\pm 1$ then correspond to the 
simultaneously measurable physical qubits of the Majorana surface code.  

It is well known that interacting Majorana plaquette models as given in Eq.~\eqref{Sysham} 
can be mapped to (planar versions of) Kitaev's celebrated toric code model
 \cite{Terhal2012,VijayFu2015,XuFu2010,Kells2014}, which 
features long-range entanglement and topological order. In fact, our perturbative
 construction of Eq.~\eqref{Sysham} directly 
mirrors Kitaev's original derivation of the toric code from a strongly anisotropic variant of his
 honeycomb model \cite{Kitaev2003}, cf.~Ref.~\cite{Terhal2012}.
We note that Refs.~\cite{Nussinov2012,Franz2015,Franz2015b,Rahmani2015,Rahmani2015b} 
discuss similar interacting Majorana networks
which may yield alternative code constructions different from a surface code.  

The mapping to the toric code also confirms that stabilizers must come 
in two different types, $\hat {\cal O}_n=\hat X$ or $\hat {\cal O}_n=\hat Z$, see Fig.~\ref{f1}.  
Indeed, by using the MCB Pauli operators ($\hat x,\hat y,\hat z)_l$ introduced above, 
we observe that away from lattice boundaries, every $\hat X$ ($\hat Z$) stabilizer
operator is equivalently expressed as product of the four $\hat x$ ($\hat z$) 
Pauli operators on the adjacent MCBs numbered by $l_1\cdots l_4$ ($l_1'\cdots l_4'$), 
\begin{equation}\label{stabilizerOp2}
\hat X=\hat x_{l_1}\hat x_{l_2}\hat x_{l_3}\hat x_{l_4},\quad
\hat Z=\hat z_{l'_1}\hat z_{l'_2}\hat z_{l'_3}\hat z_{l'_4}.
\end{equation}
The product is thus taken along the minimal path surrounding the respective plaquette.  
Again in agreement with the planar toric code, stabilizers located at the boundaries of the system 
only involve three Pauli (six Majorana) operators, see Fig.~\ref{f1}. 
It is worth noting that one could also take the fermion parity operators $\hat {\cal P}_l$, defined as the
 product of the four Majorana operators on MCB no.~$l$, as a third type of stabilizer. 
This fact suggests the possibility of a Majorana surface code equipped with three different
 types of stabilizers ($\hat X,\hat Z,\hat {\cal P})$, cf.~Refs.~\cite{VijayFu2015,VijayFu2016}.  
Under strong Coulomb blockade conditions, however, the parity constraint (\ref{parityconstraint}) 
locks all $\hat {\cal P}_l$ stabilizers into specific eigenstates. In what follows,
we therefore employ only the $(\hat X,\hat Z)$-type stabilizers as active physical qubits.

In contrast to the commonly studied toric code model with just two couplings  \cite{Kitaev2003}, 
we observe that the coupling constants ${\rm Re}(c_n)$ in Eq.~\eqref{Sysham} may effectively 
assume random values due to, e.g., the presence of tunnel phases or device imperfections. 
The temporal evolution of the system thus contains a large number of uncorrelated 
random phases $\exp[\pm i{\rm Re}(c_n) t]$ which lead to dephasing 
between corresponding stabilizer states with eigenvalues $\pm 1$.
Protection of logical qubits against such dephasing processes forms an essential 
part of QIP with the surface code and can be effected, e.g., by reducing the 
respective $t_{ll'}$ couplings through gate voltage changes, see Sec.~\ref{sec3a} below. 
Errors could also arise from higher-order corrections to the code Hamiltonian in Eq.~\eqref{Sysham}.
For $|t_{ll'}|\ll E_C$, however, all sub-leading contributions beyond 
Eq.~\eqref{Sysham} are suppressed by at least a factor $(|t_{ll'}|/E_C)^2$.  
Such terms are not expected to seriously deteriorate the operation of the code. In any case, 
they can be handled in error correction along with the aforementioned and other (more severe) error sources, see Sec.~\ref{sec4f}.

We conclude that the spin-$1/2$ variables represented by the
MCB Pauli operators \eqref{PauliMajorana} constitute the elementary hardware qubits in our setup.  
Their definition is linked to the existence of topologically protected Majorana 
states. We therefore expect that these qubits come with long coherence times
and thus represent attractive platforms for QIP operations in general \cite{paperB}.
Physical qubits (stabilizers) are then defined as products of four MCB
 Pauli operators along the minimal loops of the 2D lattice, see Eq.~\eqref{stabilizerOp2}.
For the definition of logical qubits, see Sec.~\ref{sec3} below.
 
\subsection{Access hardware}\label{sec2c}

Before moving on to the discussion of code operations, we introduce the
hardware elements needed for initialization, manipulation, and readout of qubits.
In particular, we will describe the implementation of 
tunnel conductance probes (including interference links), single-electron pumping, and the use of 
gate voltages. None of these elements involves the application of radiation fields. 
All access operations should be performed in the quasi-adiabatic limit, i.e., 
at frequency scales well below the thresholds where radiation might be produced.   
 The above elements could be integrated directly within the 2D code structure and do not 
require device technology beyond that needed to build the code itself. 
In view of the simplicity and flexibility of this architecture,
 alternative readout and/or manipulation schemes may also be implemented in an
efficient manner, see Sec.~\ref{sec4f} and Refs.~\cite{VijayFu2015,paperB}.

\subsubsection{Tunnel conductance probes}

We consider all Majorana states to be endowed with tunable tunnel 
connectors to external probe leads. 
The presence of a weak tunnel coupling $\lambda_{l}$ between a specific 
Majorana fermion $\gamma_{l}$ and the lead fermion $\Psi_l$ 
(near the corresponding tunnel contact) will be described by a tunneling operator,
see Eq.~\eqref{TunnelingOperator} with $\Psi\to \Psi_l$, $\gamma_j\to \gamma_l$,
$\hat\varphi\to \hat\varphi_l$, and $t_j\to \lambda_l$.
These connections are essential for two-terminal interferometric transport
measurements, which are our designated readout instrument for stabilizers, 
as well as for non-stabilizers like the MCB Pauli operators \eqref{PauliMajorana}.  
For the latter type of readout, we need  a tunnel link locally connecting the 
respective pair of leads.  Such links allow for additional interference terms in the 
conductance which are sensitive to non-stabilizer eigenvalues, see below and Sec.~\ref{sec4a}.

We first describe the measurement and readout of stabilizer eigenvalues.  
To that end, consider two neighboring Majorana states on adjacent MCBs, 
say, $\gamma_0$ and $\gamma_1$ in Fig.~\ref{f1}, which are
connected by the effective tunnel matrix element $t_{01}$.  
Both Majorana states are also tunnel-coupled to separate probe leads.
The application of a small bias voltage between these leads
generates current flow, and the corresponding linear conductance, $G_{01}$, 
can be computed from an effective transfer Hamiltonian $H_{01}$. 
The latter follows by lowest-order expansion in $\lambda_{0,1}$, where we also perform  
a Schrieffer-Wolff transformation in order to project to the charge ground state on each MCB, 
similar to the steps leading to Eq.~\eqref{Sysham}. We obtain 
\begin{equation}\label{TunnelAmplitude}
H_{01}=\alpha \left( \xi + c_u^\ast \hat {\cal O}_{u}+c_{d} \hat {\cal O}_{d}\right)
\Psi_{0}^\dagger \Psi_{1}^{} + {\rm h.c.},
\end{equation}
with $\alpha= - 32\lambda_{0}\lambda_{1}^*/(5 t^\ast_{01} E_C)$. 
Equation~\eqref{TunnelAmplitude} features a superposition of a constant term ($\xi$),
 describing tunneling via the direct link $\sim t_{01}$, with two subleading terms 
 of tunneling around nearby loops. The latter involve the Majorana strings $\hat{\cal O}_{u/d}$
around the upper ($u$) and lower ($d$) plaquette adjacent to the contacted pair 
$\gamma_0\gamma_1$, with the  coefficients $c_{u,d}$ in Eq.~\eqref{Sysham}.
We find $\xi=(5/16) \eta |t_{01}|^2/E_C$ with a dimensionless 
asymmetry parameter $\eta\sim \Delta n_{g,0} \Delta n_{g,1}+\mathcal{O}(\Delta n_g^4)$;  
the full expression for $\eta$ can be found in Ref.~\cite{Aviad2016}.
Importantly, $\eta$ depends on detuning parameters $\Delta n_{g,l=0,1}$
off the respective particle-hole symmetric point (with integer $n_{g,l}$) at which 
the addition/removal of one charge unit to/from the MCB costs precisely the same energy ($E_C$).
If at least one detuning vanishes, the term $\sim \xi$ will be 
absent in Eq.~\eqref{TunnelAmplitude}. This cancellation of the direct process can be traced 
back to a destructive interference between different time-orderings of tunneling events.
By varying the backgate voltage parameters $n_{g,l}$, one can then
optimize the amplitude of the interference signal between the direct 
path and the loop contributions in the conductance readout below.

With the normal density of states $\nu_{l=0,1}$ of the respective probe lead,
second-order perturbation theory in $H_{01}$, see Eq.~\eqref{TunnelAmplitude}, 
yields the tunneling conductance 
\begin{equation}\label{conductance}
\frac{G_{01}}{e^2/h} = 4\pi^2 \nu_0\nu_1 |\alpha|^2
\left( g + g_{u} {\cal O}_{u} + g_{d} {\cal O}_{d} +
g_{ud} {\cal O}_{u} {\cal O}_{d} \right).
\end{equation}
Evidently the conductance is sensitive to the stabilizer eigenvalues ${\cal O}_{u,d}=\pm 1$, where 
we tacitly assume that the conductance measurement implies a projection to the corresponding 
eigenstates, cf.~Refs.~\cite{Bonderson2007,Bonderson2008}. Apart from a stabilizer-independent contribution with
$g= \xi^2 +|c_{u}|^2+|c_{d}|^2$, the conductance \eqref{conductance} contains quantum 
interference terms between the direct link and individual loops, 
$g_{u/d}=2\xi \mathrm{Re}(c_{u/d})$, and a two-loop interference term with $g_{ud}=2\mathrm{Re}(c_{u}c_{d})$. 

The coefficients $(g, g_{u}, g_d, g_{ud})$ will show random variations between readout 
configurations $(\gamma_l,\gamma_{l'})$ targeting different pairs of stabilizers. 
However, for a given pair, they are presumably time-independent and can
 be parametrically altered by variation of the backgate parameters $n_{g,l/l'}$ and/or tunnel couplings $t_{ll'}$.
 Numerical values for these coefficients can be obtained by conductance measurements for the 
 four different plaquette configurations $(\mathcal{O}_{u},\mathcal{O}_{d})=(\pm1,\pm 1)$, where we anticipate 
 that plaquettes can be flipped in a controlled manner by pumping protocols. Once 
 the coefficients  $\{(g, g_{u},g_d, g_{ud})^{}_{ll'}\}$ have been determined for all neighbor
 pairs by this initial calibration procedure, performing a (sequential or parallelized) measurement 
 of the conductances along a set of $\approx N$ links (for a code with $N$ MCBs) 
 will determine all stabilizer eigenvalues.

Next we discuss the projective measurement and readout of non-stabilizer operators.
These basal quantum operations will be extended to logical qubits in Sec.~\ref{sec4a}, while here 
we introduce the needed hardware elements and conductance interferometry principles.
For concreteness, we consider the Pauli operator $\hat z=i\gamma_5\gamma_6$ in Fig.~\ref{f1}, with 
eigenvalues $z=\pm 1$, where a suitably designed interferometric conductance measurement 
will be sensitive to this eigenvalue, see Eq.~\eqref{conductance2} below.
 In order to perform such a measurement, one can use the external leads already 
needed for stabilizer readout. In addition, we now require an interference link locally connecting
 both leads (away from MCBs) through a tunnel matrix element $\tau$.  
 To lowest order in the tunnel couplings $\lambda_{5,6}$, transport between both terminals
is again governed by a transfer Hamiltonian ($H_{56}$).  Repeating the steps leading
 to Eq.~\eqref{TunnelAmplitude}, we now obtain 
\begin{equation}\label{TunnelAmplitude2}
H_{56}=\left( \tau +  t_z \hat z \right) \Psi_{5}^\dagger \Psi_{6}^{} + {\rm h.c.}
\end{equation}
with $t_z \sim\lambda_5^{}\lambda_6^*/E_C$.  The two-terminal conductance follows from 
perturbation theory as in Eq.~\eqref{conductance}, 
\begin{equation}\label{conductance2}
\frac{G_{56}}{e^2/h} = 4\pi^2 \nu_5\nu_6 
\left[ |\tau|^2+|t_z|^2+ 2 {\rm Re}(\tau^\ast t_z) z \right] .
\end{equation}
After an initial calibration step, a conductance measurement thus determines $z$, where the measurement again 
 implies a projection to the respective eigenstate of $\hat z$.  We note that the   
amplitude of the interference signal can be optimized by tuning the link parameter $\tau$.  
By using different lead pairs attached to the same MCB,  
one may access all Pauli operators ($\hat x, \hat y$, and $\hat z$) on this box.  
Similarly, for leads coupled to different MCBs, readout of other non-stabilizer 
operators (e.g.~the string operators below) is possible, cf.~Sec.~\ref{sec4a}.

\subsubsection{Single-electron pumping} 

Whereas conductance interferometry is our proposed readout tool, 
adiabatic charge pumping of single electrons onto and off the code by means of tunnel-coupled quantum dots
will be tailor-made for the controlled manipulation of qubits.  
For simplicity, we assume that each quantum dot can accommodate only a 
single electron (occupation number $n=0,1$), where the level energy is gate-tunable. 
When the tunnel coupling $\tilde\lambda_l$ to the corresponding Majorana 
state $\gamma_l$ is finite, by a suitable gate voltage change, one can pump (or absorb) one 
elementary charge into (from) the code \cite{Flensberg2011,Leijnse2011}.   
The tunneling Hamiltonian is given by Eq.~(\ref{TunnelingOperator}) with $\gamma_j\to \gamma_l$,
 $t_j\to\tilde\lambda_l$, $\hat\varphi\to \hat\varphi_l$, and $\Psi^\dagger\to d_l^\dagger$, 
 where $d_l^\dagger$ is the fermion operator for the dot.  Clearly, a single tunneling 
 event will drive the respective MCB outside its charge ground state, and dot operations
  categorically have to be  performed in pairs to maintain charge neutrality of the code.  
 We here assume that the pumping operation is performed in an adiabatically slow way, 
 at rates to be studied momentarily, although weak deviations from adiabaticity 
are not detrimental if combined with a measurement of the final dot occupations. 
Such control measurements collapse the system of dots into a definite state, 
thus eliminating diabatic errors that could spoil the code operation \cite{Nayak2016}.
 
 Denoting the initial charge configuration of the dot pair
 by $(n_l, n_{l'})=(1,0)$,  consider a gate-voltage protocol resulting 
 in the final state ($0,1$).  This process transfers precisely one electron 
 through the code, entering via the Majorana state $\gamma_l$ and exiting via $\gamma_{l'}$.  
The transfer Hamiltonian $H_{ll'}$ describing this process can be obtained by similar steps as those leading
 to Eq.~\eqref{TunnelAmplitude} above, where we find  
\begin{equation}\label{TunnelAmpl3}
H_{l l'} = {\cal T}_{ll'} d_l \hat {\cal S}_{ll'} d^\dagger_{l'},\quad
{\cal T}_{ll'} \simeq 
\frac{\tilde\lambda_l^\ast\tilde\lambda_{l'}}{E_C}   \prod_{j=1}^m \frac{t^\ast_{ l_j l_j' } } {E_C}  . 
\end{equation}
The transfer amplitude ${\cal T}_{ll'}$ has been specified up to a prefactor of no 
interest here, and $\hat {\cal S}_{ll'}$ represents string operators describing 
 intra-code electron transfer processes.  In general,  Eq.~\eqref{TunnelAmpl3} involves a linear superposition of
many contributions from different tunneling paths $\gamma_l\to \gamma_{l'}$ between the selected end-point Majorana states.  For an arbitrary path with $m$ tunneling
events between neighboring MCBs ($\gamma_{l_j}\to \gamma_{l'_j}$ with tunnel links $j=1,\ldots,m$), the string operator is given by 
the product of these $2m$ Majorana operators in addition to the end-point Majorana states \cite{footnotesec2},
\begin{equation} \label{stringopdef} 
\hat {\cal S}_{ll'} =  (i\gamma_l\gamma_{l'}) \hat W_{ll'},\quad \hat W_{ll'}= \prod_{j=1}^m
\left(i\gamma_{l_j}\gamma_{l_{j}'}\right).
\end{equation}  
For $|t_{l_jl'_j}|\ll E_C$, the dominant term in Eq.~\eqref{TunnelAmpl3} comes from the 
shortest path(s) with minimal number $m$ of tunnel bridges.  
To give an example, for neighboring MCBs, we have $m=1$ and the leading contribution takes the form 
$\hat W_{ll'}=i\gamma_{l_1}\gamma_{l'_1}$.  When $\gamma_l$ (and/or $\gamma_{l'}$) coincides with 
a Majorana operator in a given path $\hat{W}_{ll'}$, the destructive interference mechanism
described above for the stabilizer readout configuration in Eq.~\eqref{TunnelAmplitude} 
is activated. 
The path contribution to Eq.~\eqref{TunnelAmpl3} can then be quenched by 
tuning the backgate parameter $n_g$ ($n_g'$) on MCB no.~$l$ ($l'$) to an integer value.

Since $\hat W_{ll'}$ in Eq.~(\ref{stringopdef}) stems only
from intra-code tunneling processes $\sim \gamma_{l_j}\gamma_{l'_j}$ between neighboring MCBs,
such operators commute with all stabilizers, $[ \hat W_{ll'},\hat {\cal O}_n]=0$.  This also implies that 
any two string operators $\hat{\cal S}_{ll'}$ are equivalent up to (products of) stabilizer operators.
In a stabilized code state, the full operator content of the pumping process is therefore identified 
by the terminal Majorana operators.  In fact, the two stabilizer operators $\hat {\cal O}_n$ containing $\gamma_l$ 
are flipped by the above pumping protocol, since $\gamma_l$ anticommutes
with those two  operators but commutes with all others. Similarly, the two stabilizers that contain $\gamma_{l'}$ will be 
flipped. It is then clear that depending on the distance between $\gamma_l$ and $\gamma_{l'}$, by
such a charge transfer process one can either flip zero, two, or four stabilizers. 
The case of no stabilizer flips corresponds to the above stabilizer readout scheme, see Eq.~\eqref{TunnelAmplitude}. In accordance with the planar toric code rules \cite{Kitaev2003},
stabilizers of each type thus can only be flipped in pairs.
For instance, by pumping a single electron from $\gamma_7\to \gamma_8$ in Fig.~\ref{f1}, the two adjacent $\hat Z$ plaquettes  
will be flipped. Since both $\gamma_7$ and $\gamma_8$ belong to the
 $\hat X$ plaquette in between, this stabilizer is flipped twice and hence remains 
 invariant.    We note that each string operator \eqref{stringopdef} can alternatively be written as product
 of MCB Pauli operators along a path from $\gamma_l$ to $\gamma_{l'}$. In particular, for the 
 above pumping process, $\hat{\cal S}=i\gamma_7\gamma_8=\hat x$ 
 connects the two flipped $\hat Z$ stabilizers in Fig.~\ref{f1}.

A single-electron pumping process can be considered adiabatic if it occurs
 on times scales long against $\tau_{\rm ad}$, where the energy scale 
$\tau_{\rm ad}^{-1} = |{\cal T}_{ll'}|$ corresponds 
 to the leading-order contribution to $H_{ll'}$. 
Under these circumstances, the process will adiabatically transform an
initial state $|\Psi\rangle_{\rm initial}$ into the final state
$|\Psi\rangle_{\rm final}\sim\hat {\cal S}_{ll'}|\Psi\rangle_{\rm initial}$.
We observe that when both dots are tunnel-coupled to Majorana states 
on the same box, the pumping protocol implements elementary Pauli operations, cf.~Eq.~\eqref{PauliMajorana}.  
In Sec.~\ref{sec4e}, we will discuss powerful generalizations of the above protocols.

\subsubsection{Gate voltage adjustments} 

The application of a static voltage to electrodes in spatial proximity of code elements 
can be used to adjust several important parameters of the code. For instance, by 
tuning the MCB backgate parameters $n_{g,l}$, 
 one can alter the strength of path contributions to electron tunneling and 
 optimize the interference signal for stabilizer readout operations.
Similarly, by changing the voltage on gates near the tunnel bridges connecting 
neighboring MCBs, we may control the matrix elements $t_{ll'}$.
For example, a minimal hole can be cut into the code by severing the four 
connections of a single stabilizer loop.  

\section{Logical qubit implementation}\label{sec3}

In Kitaev's original work on the toric code \cite{Kitaev2003}, a stabilizer system 
wrapped onto a surface of toroidal geometry was shown to exhibit topological 
ground-state degeneracy $2^{2g}$, where $g$ is the genus of the surface.
The number $2g$ thus counts the unconstrained binary degrees 
of freedom. In principle, the latter are good candidates for topologically 
protected logical qubits due to their encoding into a stabilized code space.
However, toroidal surfaces are difficult to realize in terms of 2D device technology,
and the resulting qubits are almost impossible to address and/or read out.
Later on, it was realized that the torus topology can be effectively simulated in 
terms of a planar geometry \cite{Bravyi2001,Freedman2001}, thereby
paving the way for the surface code approach.

A simple example for a planar code patch with boundaries suitable to the definition of logical qubits
 is shown for our architecture in Fig.~\ref{f1}.  The shunting of Majorana 
states defines  $Z$-type (or $X$-type) boundaries, where only $\hat Z$ (or $\hat X$) stabilizers appear at the respective boundary.   
The ground state, $|\Psi\rangle$, of the code Hamiltonian \eqref{Sysham}
has to satisfy the conditions $\hat {\cal O}_n|\Psi\rangle= {\cal O}_n |\Psi\rangle$ 
with eigenvalues ${\cal O}_n={\rm sgn} [{\rm Re}(c_n)]$. 
However, these conditions are not sufficient to uniquely determine $|\Psi\rangle$. This 
can be seen by comparing the number of constraints (given by the number
of stabilizers: 17 in Fig.~\ref{f1}) to the number of independent hardware qubits  
(MCBs: 18 in Fig.~\ref{f1}).   
We conclude that for the example in Fig.~\ref{f1}, one unconstrained degree
of freedom remains. It corresponds to Pauli operators 
$\hat Z_{\rm topo}$ ($\hat X_{\rm topo}$) defined as product of all 
$\hat z$ ($\hat x$) MCB Pauli matrices along an arbitrary path connecting 
the two $Z$-type ($X$-type) code boundaries. 
Albeit ($\hat Z_{\rm topo},\hat X_{\rm topo})$ encodes a logical qubit, 
this option is not very interesting since it just yields a single qubit, 
independent of the size of the code patch.

As we discuss below, the  ground-state degeneracy of  such surface codes can be
 effectively enhanced by the controlled creation of holes.
The drilling of such a hole may be implemented in our code architecture, e.g., 
by switching off the tunnel couplings $t_{ll'}$ surrounding a minimal plaquette, see below.   
We anticipate that such a procedure will represent a trade-off game, since for a 
dense pattern of holes, erroneous flips of the encoded logical qubits will become likely.
We here introduce logical qubits in analogy to their definition in the 
bosonic surface code approach by Fowler \textit{et al.}~\cite{Fowler2012}. 
However, our Majorana surface code architecture comes with two differences of central importance.

\begin{figure*}
\centering
\includegraphics[width=13cm]{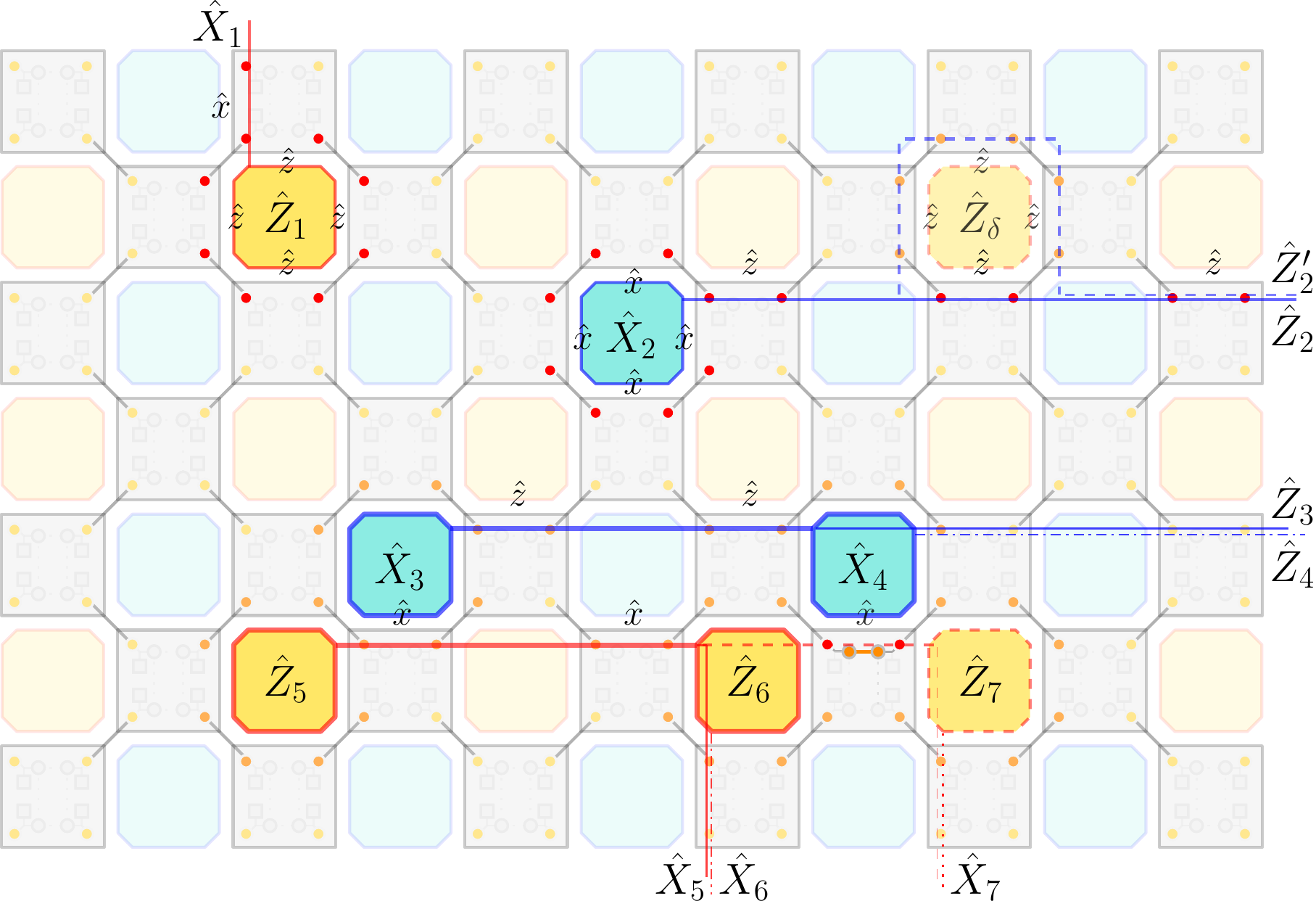}
\caption{Logical qubits in the Majorana code architecture (only a part of a 
larger 2D code is shown).  By ceasing to measure the stabilizer $\hat Z_1$ 
(yellow, top left corner), one obtains the $Z$-cut qubit ($\hat Z_1,\hat X_1)$.  
The  Pauli-$\hat X_1$ string connects $\hat Z_1$
to an $X$-type boundary (one choice is shown as red line). 
Similarly, ($\hat X_2,\hat Z_2$) is an $X$-cut qubit, where we stop measuring 
$\hat X_2$ (dark blue). The string operator $\hat Z_2$ (blue line)  connects  
$\hat X_2$ to a $Z$-type boundary, where an alternative choice $\hat
Z_2' = \hat Z_\delta \hat Z_2$ is shown as well.   
Next, the $X$-type double-cut qubit ($\hat X_A,\hat Z_A$) is
built from two single-cut qubits $(\hat X_3,\hat Z_3)$ and $(\hat X_4,\hat Z_4)$.
We use $\hat X_A=\hat X_3$ and the string operator $\hat Z_A=\hat Z_3\hat Z_4$
 connecting $\hat X_3$ and $\hat X_4$.    Similary, the $Z$-type double-cut
 qubit ($\hat Z_B, \hat X_B)$ is formed from $(\hat Z_5,\hat X_5)$ and 
 $(\hat Z_6,\hat X_6)$.  Choosing $\hat Z_B=\hat Z_5$, both plaquettes are connected by the
 internal string operator
  $\hat X_B=\hat X_5\hat X_6$.  Finally, by readout of the MCB Pauli-$\hat x$ operator
  connecting $\hat Z_6$ and $\hat Z_7$ (probe leads shown as orange circles),
   logical information can be moved from $\hat Z_6\to \hat Z_7$, i.e.,
the  double-cut qubit $B$ is then encoded between stabilizers no.~5 and 7, see Sec.~\ref{sec4b}.}
\label{f2}
\end{figure*}

First, Fowler \textit{et al.}~employ ancilla qubits (called measurement qubits in
 Ref.~\cite{Fowler2012}) in addition to the physical qubits (data qubits in Ref.~\cite{Fowler2012}). 
  By entangling four data qubits with a measurement qubit through a 
  sequence of controlled-NOT (CNOT) operations, projective stabilizer readout becomes possible.  In our case, since 
readout is performed by conductance measurements using interference loops, 
see Sec.~\ref{sec2c}, neither measurement qubits nor additional CNOT operations are necessary.
Compared to the bosonic variant, the present Majorana code architecture thus offers a significant reduction of hardware overhead 
together with a simpler single-step stabilizer measurement, see also Ref.~\cite{VijayFu2015}.

Second, Fowler \textit{et al.}~implement a logical qubit by 
stopping certain CNOT operations between measurement and data qubits \cite{Fowler2012}.  
Since our Majorana code realization does not contain measurement qubits,  
we here need to proceed in a different manner.  In fact, it is necessary 
 to implement a combined software-hardware hole.
First, to exclude a stabilizer from the cyclic readout procedure, 
the gate-tunable tunnel couplings $\lambda_l$ and $\lambda_{l'}$
between the code and the respective leads, cf.~Sec.~\ref{sec2c}, will remain turned off.  
This gives a software hole but is not yet sufficient to define a useful 
logical qubit, since the two qubit eigenstates should be (nearly) degenerate to 
avoid large dephasing errors, cf.~Sec.~\ref{sec2b}.  To achieve this goal, one may reduce the 
tunnel couplings $t_{ll'}$ by suitable gate voltage changes. The 
respective plaquette energy ${\rm Re}(c_n)$ in Eq.~\eqref{Sysham} thus decreases
and one physically cuts a hardware hole into the code.
Residual small plaquette energies can be handled in error correction, and    
to measure adjacent stabilizers one can sequentially turn on selected 
couplings $t_{ll'}$ again.

As reviewed in Ref.~\cite{Fowler2012}, one can form larger holes for implementing logical qubits
by combining several ($d$) adjacent holes. (The number $d$ is the so-called code distance 
in terms of physical qubits.) The fault-tolerance threshold theorem \cite{Terhal2015} states that as
 long as error probabilities for elementary quantum operations remain below a
 moderate threshold value of order $\approx 1\%$,  logical error rates become
 exponentially small with increasing $d$.  This exponential scaling justifies the notion of 
topologically protected logical qubits \cite{Terhal2015}.  In what follows, we 
mostly describe the case $d=1$, with the understanding that a highly efficient reduction 
of logical errors is offered by increasing $d$. All operations are implemented such that they 
directly can be adapted to the case $d>1$.

\subsection{Single-cut qubits}\label{sec3a}

Let us start with the most elementary logical qubit realizable
in our code architecture, where one forms a hole for just a single plaquette. 
According to the above discussion, the removal of 
the corresponding constraint on $|\Psi \rangle$ leaves one 
binary degree of freedom undetermined.  Since we have 
$Z$- and $X$-type stabilizers,  two different types of logical qubits are possible, and
by not measuring $\hat Z$ (or $\hat X$) we obtain a $Z$-cut (or $X$-cut) qubit, respectively.
Suppose now that we have chosen a stabilizer, say, $\hat Z_1$ in Fig.~\ref{f2}.  
By convention, we take the Pauli-$Z$ operator for this $Z$-cut qubit 
as the stabilizer operator $\hat Z_1$ itself.  However, to define a 
logical qubit, we also need to identify the conjugate Pauli operator 
$\hat X_1$ anticommuting with $\hat Z_1$. This operator should not lead out of 
the code space spanned by  simultaneous eigenstates 
of all other stabilizers.   Obviously, once $\hat X_1$ has been constructed, the 
Pauli-$Y$ operator  will also be available by virtue of the relation 
$\hat Z_1\hat X_1=i\hat Y_1$.
Below we use the shorthand notation $(\hat Z_1,\hat X_1)$ for the resulting
single-cut logical qubit.

An operator $\hat X_1$ with the required features can be constructed as 
product of all MCB Pauli-$\hat x$ operators connecting the
internal $X$-type boundary around the open $\hat Z_1$ plaquette 
to an $X$-type code boundary along an arbitrary path, see Fig.~\ref{f2}. 
Note that this construction gives $\hat X_1$ as a string operator which  
trivially commutes with all $X$-type stabilizers but also with all other $\hat Z$ stabilizers 
(apart from $\hat Z_1$).  The latter fact follows because $\hat X_1$ contains an 
even number (zero or two) of $\hat x$ operators that are conjugate to the 
$\hat z$ operators appearing in a given $\hat Z$.  The only stabilizer affected 
by the string will be $\hat Z_1$, 
with anticommutator $\{ \hat X_1,\hat Z_1\}=0$, because the terminal Pauli-$\hat x$
operator present in $\hat X_1$ anticommutes with precisely one $\hat z$ in $\hat Z_1$. 
Alternatively, switching from the spin-$1/2$ to the Majorana language, one 
can understand the above properties by noting 
that $\hat X_1$ shares a single Majorana operator with $\hat Z_1$ 
but an even number of Majoranas with all other stabilizers of the code.

Summarizing, the two operators $(\hat Z_1,\hat X_1)$ define a logical 
qubit embedded into the code space of the system. 
An $X$-cut qubit may be defined analogously by ceasing the measurement of an
$X$-type stabilizer, say, $\hat X_2$ in Fig.~\ref{f2}. With the conjugate  
Pauli operator $\hat Z_2$, which connects the $\hat X_2$-plaquette to a 
$Z$-type boundary via a string of $\hat z$ operators, we arrive at the logical qubit
$(\hat X_2,\hat Z_2)$.  

We emphasize that the definition of the string operators $\hat X_1$ and $\hat Z_2$ 
is not unique.  For instance, instead of $\hat Z_2$, equivalent choices can be written as $\hat Z'_2=\hat Z_\delta\hat Z_2$, differing from $\hat Z_2$ by an 
arbitrary product operator $\hat Z_\delta$ of $Z$-type stabilizers, see Fig.~\ref{f2} for 
an example.
However, in a stabilizer eigenstate of $\hat Z_\delta$, the difference between
$\hat Z_2$ and $\hat Z'_2$ reduces to a known sign 
corresponding to the measurement outcome $Z_\delta=\pm 1$.  
The only nontrivially related string operators in a fully stabilized code are 
those extending from the selected plaquette to different code boundaries. 
In that case, $\hat Z_2$ and $\hat Z'_2$ differ by the string operator 
  $\hat{Z}_\mathrm{topo}$ defined above. Likewise, for the $Z$-cut qubit,
   the string operator $\hat X'_1=\hat X_\delta \hat X_1$ is then related to $\hat X_1$ via 
   the operator $\hat X_\delta=\hat X_{\rm topo}$.
In a similar way, if a code patch contains multiple holes, string operators adopt 
definitions differing by the not-measured stabilizers, depending on which side they pass around the hole(s).
This property will be used extensively below in our implementation of CNOT and/or phase gates.

\subsection{Double-cut qubits}\label{sec3b}

Next we discuss how to implement a double-cut (or internal) qubit. This type of qubit
serves to disconnect logical information from the code boundary, such that all 
operations appear as for an infinite system.  In order to obtain a double-cut qubit,
two single-cut qubits may be stitched together.  For instance, consider the 
two $X$-cut qubits $(\hat X_3,\hat Z_3)$ and $(\hat X_4,\hat Z_4)$ in Fig.~\ref{f2}.
We use a redundant encoding scheme, where the two-qubit state always remains in the, say,
even-parity subspace defined by $X_3  X_4= +1$. Such a constraint can be
imposed by initial preparation and then kept throughout the computation by employing 
only a subset of qubit operators as detailed below.   
The even subspace is spanned by  $\{ |++\rangle_{34}, |--\rangle_{34}\}$,  
 where the first (second) quantum number in each ket refers to the 
eigenvalue $X_{3}=\pm 1$ ($X_4=\pm 1$) of the respective single-cut qubit.
Eigenstates of Pauli-$Z$ operators are here denoted by 
$\hat Z|0\rangle=+|0\rangle$ and $\hat Z|1\rangle=-|1\rangle$, while 
Pauli-$X$ eigenstates are  written as  
$|\pm\rangle=(|0\rangle\pm |1\rangle)/\sqrt{2}$.

We next identify $|+\rangle_A=|++\rangle_{34}$ and $|-\rangle_A=|--\rangle_{34}$ 
as the basis states of an $X$-type double-cut qubit ($\hat X_A,\hat Z_A$). 
To complete the definition of this qubit, we have to specify the 
anticommuting Pauli operators $\hat X_A$ and $\hat Z_A$.
 To that end, let us first observe that any measurement of $\hat X_3$ alone suffices 
 to stabilize the code since $X_4=X_3$ is then directly known as well.  This fact 
 suggests $\hat X_A=\hat X_3$ (or equivalently, $\hat X_A=\hat X_4$) as 
 convenient choice for the Pauli-$X$ operator.
 One easily verifies that the above $|\pm \rangle_A$ states are directly 
 eigenstates of $\hat X_A$ with eigenvalues $X_A=\pm 1$. 
The conjugate Pauli-$Z$ operator can then be chosen as 
$\hat Z_A = \hat Z_3\hat Z_4$, which anticommutes with $\hat X_A=\hat X_3$ (or 
$\hat X_4$) and gives $\hat Z_A| \pm \rangle_A=|\mp\rangle_A$. 
For plaquettes $\hat X_3$ and $\hat X_4$ located away from the 
code boundaries (as in Fig.~\ref{f2}),  $\hat Z_A$  can be  
represented by an internal string operator  not involving boundary
degrees of freedom. Indeed, all $\hat z$ operators appearing in the string $\hat Z_4$ will 
square to unity when combined with $\hat z$ operators from the same part 
of the path in $\hat Z_3$, see  Fig.~\ref{f2}.  In effect, only the $\hat z$ operators along
the internal path connecting $\hat X_3$ and $\hat X_4$ will then 
contribute to $\hat Z_A$. 

We conclude that an $X$-type double-cut qubit ($\hat X_A, \hat Z_A)$ can be obtained
by the combination of two $X$-type single-cut qubits.
 We can thereby redundantly encode an arbitrary qubit state,
 \begin{equation} \label{redundant}
 |\psi\rangle_A=\alpha|+\rangle_A+\beta|-\rangle_A,
 \end{equation}
 with complex-valued coefficients $\alpha, \beta$. In Sec.~\ref{sec4}, we describe how 
 such a state may be initialized, manipulated, and read out using the hardware elements 
 introduced in Sec.~\ref{sec2}. 
 
 By following basically the same steps as just outlined, one can
 also implement a $Z$-type double-cut qubit. 
 This is illustrated in Fig.~\ref{f2} for the qubit ($\hat Z_B,\hat X_B)$ formed from the single-cut qubits corresponding to 
 $\hat Z_5$ and $\hat Z_6$ (or $\hat Z_7$).  Furthermore, completely analogous
 arguments apply to redundant encoding schemes using the odd subspace.
 The choice regarding which subspace is employed follows from the last stabilization cycle before a pair of 
 stabilizers is left open to form the double-cut qubit. Since all proper double-cut operators work within 
 this subspace, this qubit property will be kept throughout the computation (unless errors occur which
 need to be handled in error correction).

\section{Basal quantum logical operations} \label{sec4}

So far, we have discussed the implementation of a physical qubit network and how logical qubits may be immersed into it. 
Starting from there, the realization of QIP will advance in three hierarchical stages. At the base level, we need protocols to manipulate 
logical information, that is to initialize logical qubits, to change their state, to move and braid them, and to read them out. 
These basal operations are constructed from interferometric conductance measurements and/or electron pumping protocols, 
relying on the access hardware in Sec.~\ref{sec2c}.
This will pave the way towards the second stage, the realization of a minimal set of logical gates 
for universal quantum computation. The third level implements algorithms and will not be discussed here. 

It is important to realize that the required number of basal operations increases dramatically at each consecutive level.  Let us 
briefly illustrate this point, where we also provide a short overview over this central section.  
First, given the capability to initialize and read out qubit states and/or string operators,
see Sec.~\ref{sec4a}, the movement of a logical qubit from one 
plaquette to an adjacent one will require interferometric readout, 
control measurements, and qubit flips depending on outcomes, 
see Sec.~\ref{sec4b}. Second, the formation of the braid transformation, which is 
instrumental to the realization of the workhorse gate of most QIP operations, the CNOT, then 
requires several qubit moves, see Sec.~\ref{sec4c}.  Third,
more complex operations in turn involve sequences of CNOT and/or Hadamard operations.
Furthermore, the realization of phase gates hinges on the availability of ancilla states, where we 
propose the possibility to prepare Pauli eigenstates on a separate auxiliary system, 
followed by a state injection process transferring the state to a code qubit, see Sec.~\ref{sec4d}.   
Next, in Sec.~\ref{sec4e}, we show how to prepare and manipulate qubit states in 
terms of general electron pumping protocols.  In Sec.~\ref{sec4f}, we
then offer a discussion of likely error sources and challenges in our code architecture.
While a description of quantum error correction is beyond the scope of our work, we expect that (given this
toolkit of basal operations) one can largely follow the surface code strategies 
reviewed in Refs.~\cite{Fowler2012,Terhal2015}.

In view of this proliferation of complexity, maximum efficiency and reliability 
in these basal operations will be decisive for success. 
This is, we believe, where the present approach has great advantages to offer. 
The fact that qubit measurements and state changes are realized via relatively undemanding
single-step operations as compared to, say, the five-step ancilla-state-based 
CNOT protocol proposed for bosonic codes \cite{Fowler2012},
 will have an exponential impact on the complexity of later design stages. 
Below we propose a set of basal QIP operations realizing the first development stage. 
With these operations in place, the implementation of all gates needed for universal 
quantum computation is possible as summarized in Sec.~\ref{sec5}. 
However, the essential novelties of our approach are contained in the present section.

\subsection{Initialization and projective measurements}\label{sec4a}

For the initialization of stabilizer eigenstates, we may simply take the last readout result.
In case the string eigenstate of a double-cut qubit is needed, we can utilize interferometric 
conductance measurements of MCB Pauli operators, see below and Sec.~\ref{sec2c}.
To that end, starting from a stabilized state, say, for qubit $(\hat X_A,\hat Z_A)$ in Fig.~\ref{f2}, 
one first measures all constituent Pauli-$\hat z$ operators contributing to the string operator
 $\hat Z_A$. The intermediately affected stabilizers outside of qubit $A$ then re-enter the readout cycle, 
and the $\hat Z_A$-string eigenstate (identified by outcomes $z= \pm$) remains.

Various other operations also require the initialization/readout
of MCB spin-$1/2$ states in definite Pauli ($\hat x,\hat y$, or $\hat z$) eigenstates  and/or  
the projective measurement of short string operators.   
General string operators are written as product of a few MCB Pauli (or twice as many Majorana) 
operators. Equivalently, they can be represented by (at most) two double-cut qubit 
string operators and, if multiple non-equivalent strings between two end-points are used,
a number of stabilizer operators of the surface code, cf.~Secs.~\ref{sec2c} and \ref{sec3}.
 The initialization (measurement) of MCB Pauli and/or 
string eigenstates (operators) can then be effected by conductance interferometry as 
described next.  

First, a given MCB spin can be sent into a Pauli eigenstate by connecting 
a pair of leads (equipped with an interference link) to the two  
Majorana states corresponding to the Pauli operator in question, see Eq.~\eqref{PauliMajorana}. 
 Fig.~\ref{f2} illustrates such a measurement for the Pauli-$\hat x$ operator connecting 
the stabilizers $\hat Z_6$ and $\hat Z_7$. 
The conductance then contains an interference contribution sensitive to the measured Pauli 
eigenvalue, see Eq.~\eqref{conductance2}, and the measurement projects the system 
to the corresponding eigenstate.  In case that the opposite eigenstate is needed, 
one can employ a single-electron pumping process (see Sec.~\ref{sec2c}) to flip to 
the other eigenvalue.  Alternatively, if only so-called Clifford operations (mapping
Pauli to Pauli operators) are performed, such flips can be processed by classical 
software, see Ref.~\cite{Fowler2012} for details.

\begin{figure}
\centering
\includegraphics[width=7cm]{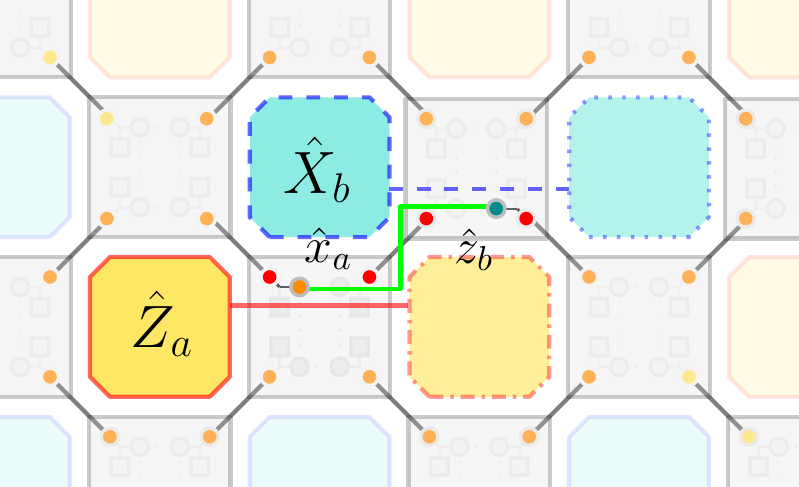}
\caption{Similar to Fig.~\ref{f2} but illustrating string operator 
measurements, where $\hat{\cal S}=\hat x_a\hat z_b$
is measured by conductance interferometry using the marked leads (orange and blue circles) sharing an 
interference link (green).  Noting the equivalent representation $\hat {\cal S}=\hat X_a\hat Z_b$, 
an arbitrary state $|\psi\rangle$ can be moved from the $Z$-type double-cut 
qubit $(\hat Z_a,\hat X_a)$ to the adjacent $X$-type qubit $(\hat X_b,\hat Z_b)$. 
An alternative approach employs the stabilizer 
product operator $\hat{\cal M}=\hat Z_a\hat X_b$, see main text. }
\label{f3}
\end{figure}

Next we describe how to measure string operators.  As concrete example, an 
implementation of the Hadamard gate is possible by measuring the string 
operator $\hat {\cal S}= \hat x_a \hat z_b=\hat X_a\hat Z_b$ in Fig.~\ref{f3}, see 
Secs.~\ref{sec4b} and \ref{sec5b} below.
Noting that $\hat {\cal S}$  affords an equivalent representation as a product of four Majorana operators, one can access 
$\hat{\cal S}$-eigenstates by activating the leads connected to the outermost of these four Majoranas, see Fig.~\ref{f3}. 
For leads sharing an interference link, in close analogy to Eq.~\eqref{conductance2}, the conductance will then 
depend on the eigenvalue ${\cal S}=\pm 1$.  This interferometric measurement also implies a 
projection to the corresponding string eigenstate. 

An alternative measurement offering similar functionality can be employed for the stabilizer product 
operator $\hat{\cal M} = \hat Z_a\hat X_b$. To this end, one may activate the tunnel links 
surrounding $\hat Z_a$ and $\hat X_b$ in Fig.~\ref{f3}, while quenching the one shared by both stabilizers. 
Projection onto individual eigenstates is thus avoided, and an adjacent pair of leads in stabilizer readout configuration (cf.~Sec.~\ref{sec2c}) 
may access ${\cal M} = \pm 1$. In this intra-code interferometry approach, no interference links are required beyond those already
needed for MCB Pauli operator readout.

\subsection{Moving logical qubits}\label{sec4b}

Any gate implementation involves the transfer of logical qubits, either as part of
the process itself, or to move information to some desired storage zone
afterwards. In the following, we first discuss a single-cell shift 
where the target qubit is of the same type as the source qubit, 
followed by the description of how to perform a state transfer between    
 adjacent qubits of different ($Z/X$) type.
Extended motion paths can then be realized by iteration.

\subsubsection{Same-type qubits}

We start by describing the state transfer between two same-type qubits.
For concreteness, let us consider the single-cut qubit $(\hat Z_6,\hat X_6)$ in
Fig.~\ref{f2}, assumed to be prepared in an arbitrary state $|\psi\rangle$. 
We wish to move $|\psi\rangle$ from this source qubit $a$, with $\hat Z_a=\hat Z_6$, 
to the adjacent target qubit $b$, given by $(\hat Z_7,\hat X_7$) in Fig.~\ref{f2}, with $\hat Z_b=\hat Z_7$. 
From the last stabilization cycle, the target qubit has been prepared in a known 
stabilizer state, say, the Pauli-$Z$ eigenstate $|0\rangle_b$ with eigenvalue $Z_b = +1$. 
We thus start from the product initial state
\begin{equation}\label{move0}
|\Psi_0\rangle =|\psi\rangle_a \otimes |0\rangle_b  = (\alpha|0\rangle +\beta|1\rangle)_a\otimes |0\rangle_b,
\end{equation}
where all operations are designed such that only qubits explicitly addressed during the move 
are affected and all other physical qubit states can be factored out.  
When the move is completed, $\hat Z_a$ has become part of the stabilizer readout cycle,
while  $\hat Z_b$ is left open and then contains the state $|\psi\rangle$.  We note that there is no need to
specify where the string operators $\hat X_a$ and $\hat X_b$ terminate. In fact, our
protocol will directly apply to a double-cut qubit, say, $(\hat Z_B,\hat X_B)$ in
Fig.~\ref{f2} with both strings ending at $\hat Z_5$. 
  
In order to move the logical information, one needs to perform a projective measurement that 
entangles the source and target qubits.  Such a measurement can be implemented by readout of 
the MCB Pauli operator $\hat x_{ab} = \hat X_a\hat X_b$, cf.~also Fig.~\ref{f2}, 
which we have already described in Secs.~\ref{sec2c} and \ref{sec4a}.
Note that $\hat x_{ab}$ is precisely the string operator connecting  $\hat Z_a$ and $\hat Z_b$. 
Assuming the measurement outcome $x_{ab} = +1$, the state $|\Psi_1\rangle$ found after the readout 
is projected to the corresponding eigenstate of $\hat x_{ab}$.
(For the other outcome $x_{ab}=-1$, see below.)  
Switching to the Pauli-$X$ basis and taking into account $\langle \hat X_a \hat X_b\rangle=x_{ab}=+1$, 
we find the (not normalized) state
\begin{equation}\label{move1}
|\Psi_1\rangle = (\alpha+\beta) |++\rangle_{ab} +(\alpha-\beta)|--\rangle_{ab}.
\end{equation}
For generic $\alpha,\beta$,  this is an entangled state of the source and target qubits.

We now include $\hat Z_a$ in the next stabilizer readout cycle (assuming outcome $Z_a=+1$)
but cease to measure  $\hat Z_b$.  Switching back to the Pauli-$Z$ basis and projecting
 the source qubit to $|0\rangle_a$, we arrive at 
\begin{equation}\label{move2}
|\Psi_2\rangle = |0\rangle_a\otimes \left(\alpha|0\rangle +\beta|1\rangle\right)_b=
|0\rangle_a\otimes |\psi\rangle_b.
\end{equation}
Evidently, we have achieved a perfect transfer of the state $|\psi\rangle$ from the source to the target qubit. 
 
In the above discussion, we have made three assumptions about measurement
outcomes, namely $Z_b=+1$ for the initial, $x_{ab}=+1$ for the intermediate, 
and $Z_a=+1$ for the final readout.  However, different outcomes 
can be accomodated by applying suitable Pauli-flip operators to the target qubit.
In particular, the state $|\Psi_2\rangle$ in Eq.~\eqref{move2} can be reconstructed from the 
bare final state $|\Psi_2\rangle_{\rm bare}$ obtained from the above protocol,
\begin{equation}\label{finalmove}
\left|\Psi_2\right\rangle = \left( \hat Z_b \right)_{x_{ab}=-1} 
\left( \hat X_b\right)_{Z_a Z_b=-1} \left|\Psi_2\right\rangle_{\rm bare},
\end{equation}
where the first (second) Pauli operator is only applied for $x_{ab}=-1$ (for $Z_a Z_b=-1$).
For Clifford-only operations, as mentioned above, the Pauli flips in Eq.~\eqref{finalmove} can 
be taken into account by the classical software used for processing measurement outcomes.
We conclude that this protocol for moving quantum information by one unit
cell between same-type stabilizers can be employed for arbitrary outcomes. 
 
As with any unitary quantum operation, such a qubit move affords either
a Schr\"odinger interpretation, in which the state changes as $ |\psi\rangle_a\otimes |0\rangle_{b}\to |0\rangle_a\otimes |\psi\rangle_{b}$, 
 or a Heisenberg view, in which states remain invariant 
but the logical qubit operators change as $(\hat Z,\hat X)_a\to (\hat Z,\hat X)_b$. 
The picture in which operators, rather than states, are subject to manipulation will often be
more convenient in the description of logical qubit operations below.

\subsubsection{Different-type qubits}

Finally, by a modified version of the above protocol, we can also transfer a state $|\psi\rangle$ 
between different ($Z/X$) types of logical qubits, see Fig.~\ref{f3}.  
In that case, we start from an initial two-qubit state 
\begin{equation}
|\Psi_0\rangle= (\alpha |0\rangle +\beta|1\rangle)_a\otimes |+\rangle_b=|\psi\rangle_a\otimes|+\rangle_b,
\end{equation}
where the state $|\psi\rangle$ is stored on the $Z$-type qubit $a$, and the $X$-type target qubit $b$ has been prepared in the $X_b=+$ eigenstate, see
 Fig.~\ref{f3}. 
Instead of a single MCB Pauli operator, in the present case we have to 
measure the string operator $\hat {\cal S}=\hat x_a\hat z_b$ in Fig.~\ref{f3}
in order to perform the move operation. Note that $\hat  {\cal S}$ is equivalent to the product of the internal 
 string operators of both qubits, $\hat {\cal S}=\hat X_a\hat Z_b$. 
This measurement can be done as described in Sec.~\ref{sec4a}, and we momentarily
assume  the outcome ${\cal S}=+1$. 

To show that the $\hat{\cal S}$-measurement indeed effects the state transfer from $a$ to $b$, 
we observe that in analogy to Eq.~\eqref{move1}, the projection to ${\cal S}=+1$ yields the intermediate state
\begin{equation}
|\Psi_1\rangle= (\alpha+\beta) |+\rangle_a\otimes|0\rangle_{b} + (\alpha-\beta)|-\rangle_a\otimes |1\rangle_{b}.
\end{equation}
Letting qubit $a$ enter the stabilizer readout cycle and assuming the outcome $Z_a=+1$,
we arrive at the final configuration 
\begin{equation}
|\Psi_2\rangle= |0\rangle_a \otimes (\alpha |+\rangle +\beta|-\rangle)_b= |0\rangle_a \otimes (\hat H|\psi\rangle)_b.
\end{equation} 
Recalling that the Hadamard ($\hat H$) operation effectively exchanges $\hat X$ and $\hat Z$ operators \cite{Nielsen2000},
the final state of qubit $b$ is interpreted as the target state of an $\hat H$-operation applied to the 
initial state $|\psi\rangle$ of qubit $a$.  
For other measurement outcomes, one again has to apply Pauli-flip operators (or classical post-processing) to qubit $b$.

Finally, a variant of this protocol can be implemented  by using the stabilizer product operator $\hat{\cal M} = \hat Z_a\hat X_b$. 
After initial preparation of a string eigenstate $Z_b = \pm 1$ and entangling qubits $a$ and $b$ via measurement of $\hat{\cal M}$, one then 
obtains the Hadamard-transformed state on qubit $b$ by measurement of the string operator $\hat X_a$. 

We conclude that the state $|\psi\rangle$ initially stored on the $\hat Z_a$ stabilizer of qubit $a$ has been 
transferred to storage in the $\hat X_b$ stabilizer of qubit $b$. Since they are of different ($Z/X$) type, this transfer effects 
a Hadamard gate on the logical state, cf.~Sec.~\ref{sec5b} below.

\subsection{Braiding logical qubits}\label{sec4c}

\begin{figure}
\centering
\includegraphics[width=7.5cm]{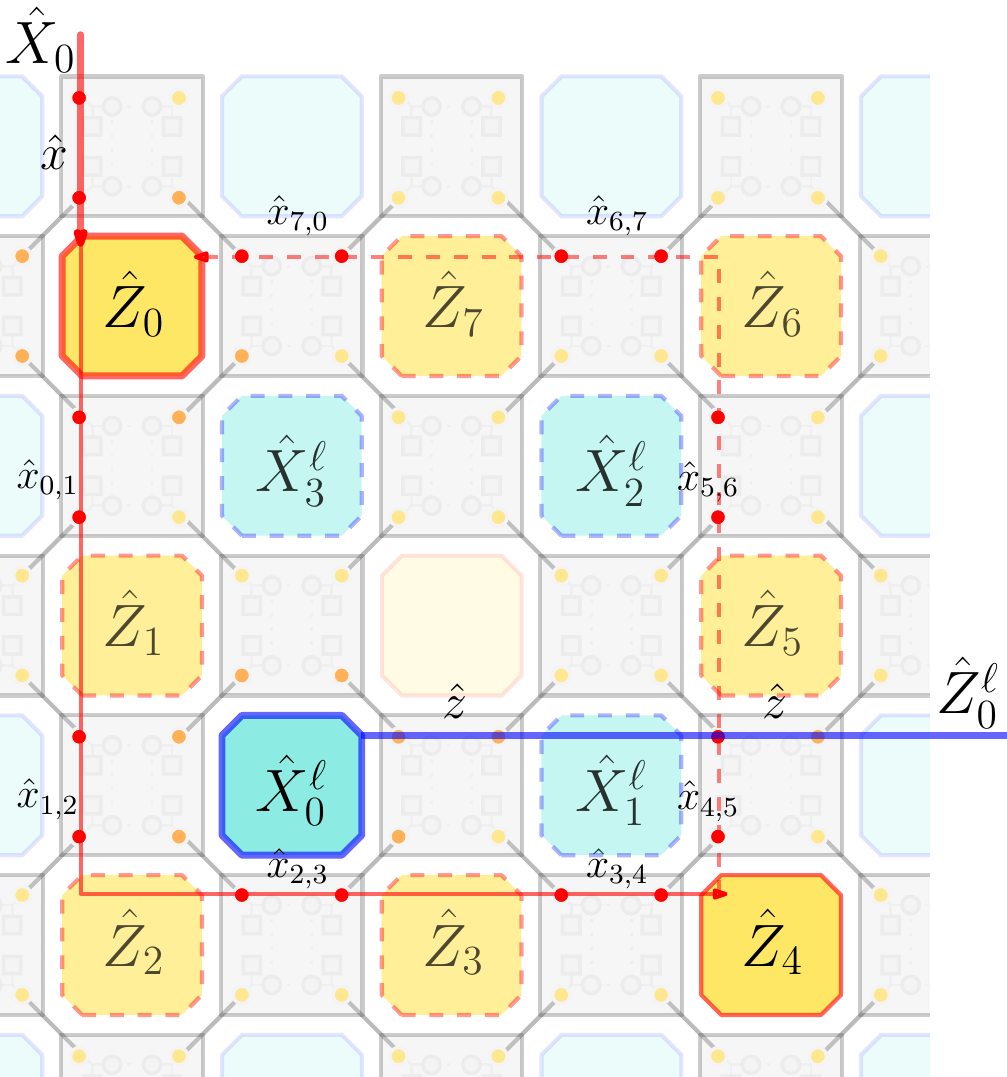}
\caption{Similar to Fig.~\ref{f2}, now for a braid operation of logical 
qubits  $(\hat Z_A,\hat X_A)=(\hat Z_0,\hat X_0$) and  $(\hat X_B,\hat Z_B)=(\hat X_0^\ell,\hat Z_0^\ell)$. 
The stabilizers $\hat Z_0$ (yellow) and $\hat X_0^\ell$ (blue) are highlighted, and
a choice for the string operators $\hat X_0$ (thick red line) and $\hat Z_0^\ell$ (blue solid line) is shown. 
By iterating the single-step operations in Sec.~\ref{sec4b} (or by analogous multi-step
moves), qubit $A$ is moved to $\hat Z_A= \hat Z_4$ and $\hat X_A=\hat X_4$.
We then bring $A$ back to the initial position by encircling $B$. After a full loop, one
has crossed the string $\hat Z_B$. This braid transformation entangles both qubits according to 
Eq.~\eqref{braid-op} and thereby implements a CNOT gate.
\label{f4}}
\end{figure}

The motion of logical qubits leaves information unchanged unless the strings forming 
part of the construction of cut-qubits  are crossed.   By contrast, when moving one qubit around another one,
such a crossing must necessarily occur.  Using the same convention for string definitions (cf.~Sec.~\ref{sec3})
before and after this qubit braid transformation, the crossing enforces a transformation of qubit operators 
equivalent to the fundamental CNOT operation, see below and Sec.~\ref{sec5a}.   
In the following we illustrate this concept for the basic braid operation of 
transferring a $Z$-type qubit around an $X$-type qubit. 

To this end, consider $(\hat Z_A,\hat X_A)$ and $(\hat X_B,\hat Z_B)$ in Fig.~\ref{f4} with the
shown convention for string operators.  We aim to  move  $A$ along a closed loop encircling $B$. 
This operation affects $B$ because in the process, the $\hat Z_B$-string emanating from the $\hat X_B$ plaquette 
 must be crossed, i.e., $(\hat Z_{A/B}, \hat X_{A/B})_\mathrm{initial}\to (\hat Z_{A/B},
\hat X_{A/B})_\mathrm{final}$ changes both qubits.
To understand how, imagine $\hat Z_A$ moved along the loop in Fig.~\ref{f4}, 
for instance by a sequence of single-step movements as described in Sec.~\ref{sec4b}.  
Up to a known sign, this brings $(\hat Z_A)_\mathrm{final}= 
(\hat Z_A)_\mathrm{initial}$ back to the original configuration.  We stress that the same initial and final conventions
are taken to define the string operators $\hat X_A$ and $\hat Z_B$, which escape directly towards the top and the right boundary in Fig.~\ref{f4}, respectively.
The string operator $\hat X_A$ will then change 
as $(\hat X_A)_\mathrm{final}=\hat X_\mathrm{loop}(\hat X_A)_\mathrm{initial}$,  where the loop operator
\begin{equation}\label{Xloop}
\hat X_{\rm loop}=\hat x_{0,1}\hat x_{1,2}\cdots\hat x_{7,0}
\end{equation} 
comprises the product of all MCB Pauli operators $\hat x_{l,l+1}$ adding to the string 
along its stepwise extension around the loop in Fig.~\ref{f4}.  

The operator  \eqref{Xloop} affords the alternative representation
\begin{equation}\label{Xloop2}
\hat X_\mathrm{loop}= \hat X_0^{\ell} \hat X_1^\ell \hat X_2^\ell \hat X_3^\ell,
\end{equation}
where $\hat X_j^\ell$ denotes the $\hat X$ operators tiling the area 
surrounded by the loop.
The equivalence of both representations follows from the fact that of the $4\times 4$ Pauli-$\hat x$ operators entering the product of $\hat X_j^\ell$ operators in 
Eq.~\eqref{Xloop2}, 
all those sitting on links in the bulk of the area will pairwise square out. 
The uncompensated $\hat x$ at the loop boundary are precisely those appearing
in Eq.~\eqref{Xloop}. 
Now, of the four $\hat X_j^\mathrm{\ell}$ stabilizers appearing in the product \eqref{Xloop2}, three are part of the readout cycle 
and thus only contribute a 
known sign to the result. The only exception is $\hat X_0^\mathrm{\ell}=\hat X_B$,  and (up to that sign) we arrive at
$(\hat X_A)_\mathrm{final}=(\hat X_B)_\mathrm{initial}(\hat X_A)_\mathrm{initial}$. 

Finally, the change of $\hat Z_B$ follows from the interpretation in which 
the motion of  $A$ around  $B$ is viewed as one of $B$ around $A$. This amounts to a crossing of the $\hat Z_B$-string with the $\hat X_A$-string,
 and by the argument above, we have $(\hat Z_B)_\mathrm{final}=(\hat Z_A)_\mathrm{initial}(\hat Z_B)_\mathrm{initial}$.  At the same time, $\hat X_B$ remains unchanged in the process, $(\hat X_B)_\mathrm{final}=(\hat X_B)_\mathrm{initial}$. 
 
 These braiding rules are summarized by the four basic transformation laws  (with identity $\hat I$)
\begin{eqnarray} \label{braid-op}
\hat Z_A\otimes \hat I_B &\to& \hat Z_A\otimes \hat I_B,\quad \hat I_A\otimes \hat Z_B\to \hat Z_A\otimes \hat Z_B,\\
\nonumber
\hat X_A\otimes \hat I_B&\to & \hat X_A\otimes \hat X_B,\quad \hat I_A\otimes \hat X_B\to \hat I_A\otimes \hat X_B.
\end{eqnarray}
The braid operation is of paramount importance in that it implements the
CNOT operation on the two qubits $A$ and $B$, see Sec.~\ref{sec5a}.

\subsection{State injection} \label{sec4d}

\begin{figure}
\centering
\includegraphics[width=6cm]{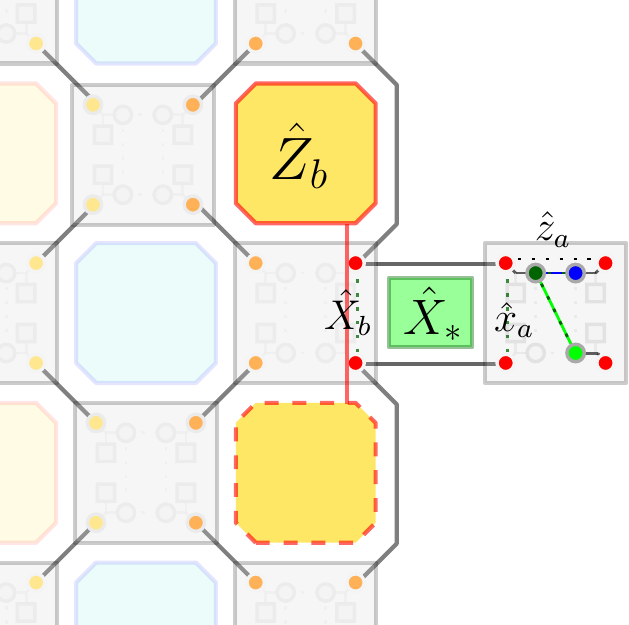}
\caption{Setup for state injection into the surface code. 
An auxiliary MCB with Pauli operators $(\hat x_a,\hat y_a,\hat z_a)$ 
is tunnel-coupled to a MCB belonging to the code but located 
near a ($Z$-type) boundary.  An arbitrary state $|\psi\rangle$ may now be prepared on 
the decoupled MCB $a$.  Pauli eigenstates can be initialized by conductance interferometry, 
where the green dotted line indicates a lead pair (with interference link) for measuring $\hat y_a$.
Subsequently connecting MCB $a$ to the code, ring exchange processes identify the
operator $\hat X_\ast=\hat X_b\hat x_a$ (green square), where $\hat X_b$ is the internal
string operator for the double-cut code qubit $(\hat Z_b,\hat X_b)$. 
By measuring $\hat X_\ast$ via conductance interferometry, 
and finally disentangling MCB $a$ from the code (by measurement of $\hat z_a$), 
$|\psi\rangle$ is transferred to the target qubit $b$.  }
\label{f5}
\end{figure}

It may often be convenient to initialize a state in a separate auxiliary system, where the 
state is subsequently injected into a target logical qubit near the boundary of 
the bulk surface code. A setup allowing for such a state injection process is shown in Fig.~\ref{f5}, where an
auxiliary MCB with Pauli operators $(\hat x_a,\hat y_a,\hat z_a)$ is tunnel-coupled 
to a MCB within the code but located at its boundary.  This process can be particularly
helpful for generating the ancilla states needed for phase gate implementations (see Sec.~\ref{sec5}) 
since the state preparation can be optimized on a separate device away from the code.  

Initially, the auxiliary MCB  $a$ is assumed to be decoupled, i.e., tunnel couplings 
to the code are turned off.  We then imagine preparing an arbitrary state 
$|\psi\rangle=\alpha|0\rangle+\beta |1\rangle$ on this MCB, without affecting the 
decoupled code state.  For instance, Pauli eigenstates can be directly initialized
 as discussed in Sec.~\ref{sec4a}. 
 As target qubit, we consider the double-cut qubit ($\hat Z_b,\hat X_b$) near the code boundary in Fig.~\ref{f5}. 
Assuming the initialization $Z_b = +$, we thus start from 
\begin{equation}
|\Psi_0\rangle = |\psi\rangle_a\otimes |0\rangle_b.
\end{equation}  
We next switch on the tunnel couplings connecting MCB $a$ to the code,
and cease to measure the two $\hat Z$-stabilizers defining the target qubit.
To achieve the state transfer from $a$ to $b$, we then perform a projective 
measurement of the operator $\hat X_\ast=\hat X_b\hat x_a$ (green square in Fig.~\ref{f5}), 
which anticommutes with both $\hat Z_b$ and $\hat z_a$. This can be done 
by conductance interferometry using a pair of leads attached to (say) the upper 
two Majorana states in the $\hat X_\ast$-loop, similar to the stabilizer readout
described in Sec.~\ref{sec2c}.   

We can verify the state transfer from MCB $a$ to the code qubit $b$ as follows.
By writing $|\Psi_0\rangle$ in the Pauli-$X$ basis for both qubits 
and then projecting to the $\hat X_\ast$-eigenstate with, say, $X_\ast = +$, we obtain the intermediate state
\begin{equation}
|\Psi_1\rangle = (\alpha+\beta)|++\rangle_{ab}+(\alpha-\beta)|--\rangle_{ab}.
\end{equation}
We now disentangle both qubits by measuring $\hat z_a$ on MCB $a$ via 
conductance interferometry. Assuming the outcome $z_a = +$, we arrive at the final state 
\begin{equation}
|\Psi_2\rangle = |0\rangle_a\otimes (\alpha|0\rangle_b+\beta|1\rangle_b) = |0\rangle_a\otimes |\psi\rangle_b.
\end{equation}
For other outcomes, one may have to apply Pauli flip operators as described for a qubit move in Sec.~\ref{sec4b}.

We have thereby injected an arbitrary ancilla state $|\psi\rangle$ from the
auxiliary MCB (where it has been prepared) to a target qubit of the 
surface code.  Another injection step can then take place as soon as the 
logical information on the target qubit has been transported away from the code boundary.

\subsection{Multi-step pumping protocols} \label{sec4e}

A major challenge for any universal quantum computation scheme is to implement 
a non-Clifford gate such as the general phase gate $\hat P(\theta)$, where the single-qubit operator
\begin{equation}\label{phasegateop}
\hat P(\theta)= e^{i\theta \hat Z}
\end{equation}
performs a Bloch sphere rotation around the $Z$-axis. In particular, the $\hat T$-gate (also called $\pi/8$-gate) follows by
choosing the phase angle $\theta = -\pi/8$ \cite{Nielsen2000}.  
As discussed in Sec.~\ref{sec5d} below, it may be advantageous to realize the phase gate operation on 
an arbitrary logical qubit state $|\psi\rangle$ by first generating ancilla states using $\hat P(\theta)$. 
Such a state can then be consumed in a quantum circuit that entangles the logical and ancilla qubit 
during intermediate steps of the protocol, where the (perfect copy of an) initial ancilla state is given by
\begin{equation}\label{ancilla-theta}
|A_\theta\rangle = \hat P(\theta)|+\rangle = (e^{i\theta}|0\rangle + e^{-i\theta}|1\rangle)/\sqrt2~.
\end{equation}
For a successful phase gate implementation, we thus
have to be able to accurately 
prepare this specific ancilla state. In what follows, we show that this task can be achieved in our 
code architecture by means of a four-step adiabatic electron pumping protocol.
The latter can be experimentally realized by sweeping gate voltages on  
quantum dots which in turn are tunnel-coupled to individual Majorana states, 
see Sec.~\ref{sec2c}.
 
\begin{figure}
\centering
\includegraphics[width=7.5cm]{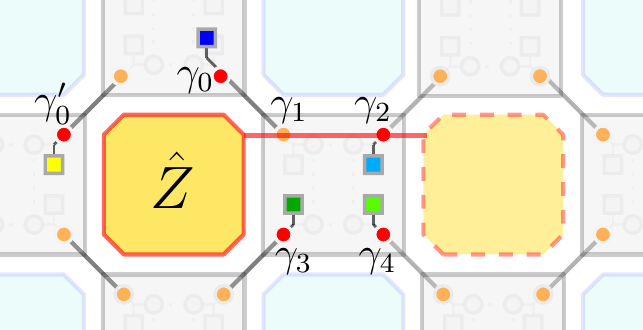}\\

\vspace*{0.6cm}
\includegraphics[width=6cm]{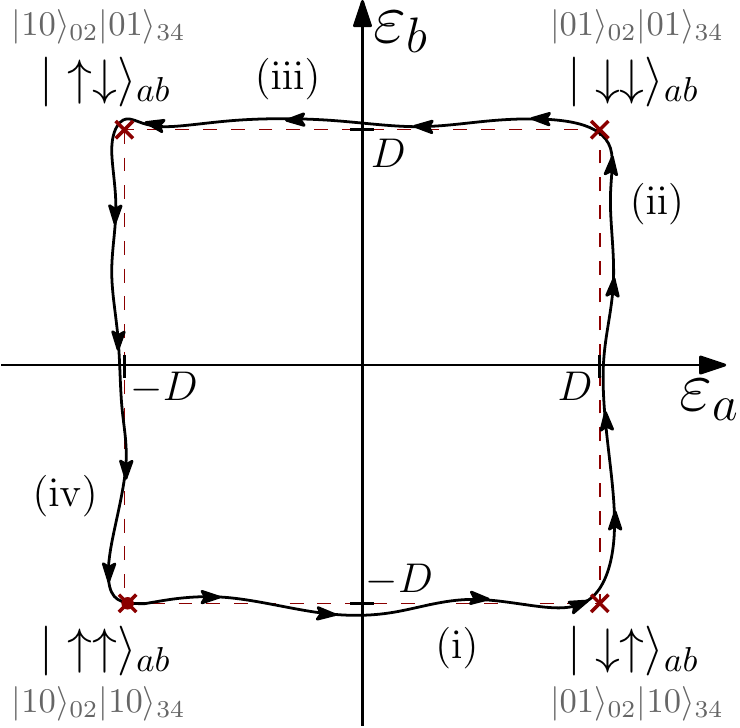}
\caption{Upper panel: Phase gate implementation for a $Z$-type double-cut qubit ($\hat Z,\hat X$),
initialized in the eigenstate $|+\rangle$ of the internal string operator $\hat X=i\gamma_1\gamma_2$. 
Two pairs of single-level quantum dots 
(dark/light blue and dark/light green squares connected to $\gamma_{0/2}$ and $\gamma_{3/4}$, respectively) 
are activated to pump single electrons through the code. After running a four-step pumping protocol
the final qubit state is $|A_\theta  \rangle=\hat P(\theta)|+\rangle$, 
where the phase gate operator $\hat P(\theta)$ has tunable angle $\theta$, see Eq.~\eqref{thetavar}.  
By using the alternative dot configuration $0\to 0'$ (yellow square),
parametrically different angles $\theta$ can be generated.
Lower panel: Schematic illustration of the four-step protocol (\ref{protocol1}) obtained 
by sweeping the dot energy levels $\varepsilon_{a/b}$, see Eq.~\eqref{energylevels}, via gate voltages. 
Intermediate states of the auxiliary ($a,b$) spin variables are also specified in the physical dot occupation number 
representation. Straight dashed lines show idealized protocols, where $\varepsilon_a(t)$ during step (iii) is a time-reversed copy of 
step (i), and similarly for $\varepsilon_b$ during steps (ii,iv).  The solid curve is for a realistic (imperfect)  protocol, see main text.
}
\label{f6}
\end{figure}

As ancilla qubit, we consider the $Z$-type double-cut qubit $(\hat Z, \hat X)$ in Fig.~\ref{f6}. 
By means of an interferometric measurement, see Sec.~\ref{sec4a}, the qubit is assumed to be
initially prepared in the $|+\rangle$-eigenstate of the internal string operator $\hat X$. 
We show below that the state $|A_\theta\rangle$ in Eq.~\eqref{ancilla-theta} can then be generated by  
a multi-step pumping protocol employing two pairs of single-level quantum dots denoted by (0,2) and (3,4) 
in Fig.~\ref{f6}.  Each dot has a gate-tunable level energy $\varepsilon_j$ and is attached to the respective Majorana 
operator $\gamma_j$ (with $j=0,2,3,4$). 
The system will be operated in a parameter regime where each dot pair holds a single electron.
For simplicity, we consider antisymmetric energy configurations, where only the two energies $\varepsilon_{a/b}$ with 
\begin{equation}\label{energylevels}
\varepsilon_a/2 = \varepsilon_0 = -\varepsilon_2, \quad \varepsilon_b/2 = \varepsilon_3=-\varepsilon_4 ,
\end{equation}
are changed during the protocol.  With the large positive energy scale $D$, by slowly varying $\varepsilon_a$ 
from $-D \to +D$ (or in reverse order $+D\to - D$), a single electron will be adiabatically pumped from dot $0\to 2$ (or vice versa).
By sweeping $\varepsilon_b$, we can similarly pump an electron between dots 3 and 4.

Let us first address the dot pair $a=(0,2)$ with fermion operators $(d_0,d_2)$ subject to 
the condition $d^\dagger_0 d^{}_0 +d^\dagger_2 d^{}_2=1$.  It is instructive to describe this dot pair 
by the equivalent spin-$1/2$ operator $\vec S_a=(S_a^x,S_a^y,S_a^z$) with
\begin{equation}\label{spinadef}
S_a^z = \frac12(d_0^\dagger d_0^{} -d_2^\dagger d_2^{}), \quad S_a^+ = d_0^\dagger d_2^{}, \quad S_a^-=(S_a^+)^\dagger,
\end{equation}
where $S_a^\pm=S_a^x\pm i S_a^y$.  We denote the eigenstate of $S_a^z$ to eigenvalue $\pm 1/2$ as 
$|\uparrow/\downarrow\rangle_a$. The corresponding occupation number states on this dot pair are 
 $|10\rangle_{02}=|\uparrow\rangle_a$ and $|01\rangle_{02}=|\downarrow\rangle_a$, see Fig.~\ref{f6}.
 By steps as detailed in Sec.~\ref{sec2c}, after a Schrieffer-Wolff projection 
to the charge ground state of the MCBs, transfer processes between dots ($0,2$) are then encoded by 
the effective spin-$1/2$ Hamiltonian
\begin{equation}\label{HpumpA}
H_a  = \varepsilon_a S_a^z +  \hat{T}_a S_a^+ + \hat T_a^\dagger S_a^-,\quad 
\hat{T}_a = it^x_a\hat X +t^y_a \hat Y. 
\end{equation}
The transfer operator $\hat{T}_a$ comes from summing over all possible tunneling paths through the code which connect the two dots. 
We here assume that physical qubits surrounding the double-cut qubit in Fig.~\ref{f6} are either in a stabilized state or turned off 
(see Sec.~\ref{sec2c}). In either case, there are only two different path contributions 
to $\hat T_a$, where the complex-valued amplitudes $t^x_a$ and $t_a^y$ depend on the 
intra-code tunnel couplings $t_{ll'}$ along the respective paths. They also contain an overall prefactor 
due to the dot-Majorana tunnel couplings, cf.~Eq.~\eqref{TunnelAmpl3}.

The first path $\sim t_a^x$ proceeds along the direct $\gamma_0\to \gamma_1$ link in Fig.~\ref{f6}.  Since it
includes in- and out-tunneling events via $\gamma_0$,   the Majorana operator $\gamma_0$
appears twice and thus squares out. Hence this contribution to $\hat T_a$ involves the internal string 
operator $\hat X= i\gamma_1 \gamma_2$, see Eq.~\eqref{HpumpA}.  
The second path involves an additional loop around the stabilizer $\hat Z$, which generates the Pauli operator 
$\hat Y=-i\hat Z\hat X$ and thereby explains the term $\sim t^y_a\hat Y$ in Eq.~\eqref{HpumpA}.
Because of the destructive interference mechanism discussed in Sec.~\ref{sec2c},
we obtain $t^x_a=0$ by choosing an integer backgate parameter $n_g$ (detuning $\Delta n_g \to 0$) for the MCB hosting $\gamma_0$.
Since $t_a^y$ involves more tunneling events than $t_a^x$, we can then tune the relative strength of both paths over a wide range, 
$|t_a^x/t_a^y|\approx \Delta n_g E_C^2/\bar t^2$, where $\bar t$ denotes a typical value of $|t_{ll'}|$.

Similarly, processes employing the dot pair $b=(3,4)$ can be described in the spin language. 
Defining spin-$1/2$ operators $\vec S_b$ as in Eq.~\eqref{spinadef} but with $d_0\to d_3$ and $d_2\to d_4$,
 the effective spin Hamiltonian $H_b$ is as in Eq.~\eqref{HpumpA} with $a \to b$.
However, the respective coupling ratio $|t^y_b/t^x_b|$ is now much smaller, 
since the longer path encircling the $\hat Z$ plaquette is both subject to destructive interference and
suppressed by orders in $|t_{ll'}|/E_C$, see Fig.~\ref{f6}.

Using a four-step shuffling sequence alternating between dot pairs $a$ and $b$, single electrons are then
transferred back and forth between the two dots in a given pair.  This forced transfer process
employs interfering tunneling paths running through the code itself and will allow us to manipulate 
the logical qubit state in a controllable manner. In fact, gate voltage sweeps of $\varepsilon_{a/b}$ are 
equivalent to independent magnetic field changes, where the field is always oriented along the respective $z$-axis and 
may induce spin flips $|\uparrow\rangle_{a/b} \to |\downarrow\rangle_{a/b}$ or vice versa.
The two spins $\vec S_{a,b}$ in turn interact with the code qubit $(\hat Z,\hat X)$, see Eq.~\eqref{HpumpA},
resulting in a gate-voltage steered motion of the qubit state on the corresponding Bloch sphere.  
Our protocol starts out from the initial spin configuration $|\uparrow\uparrow\rangle_{ab}$, corresponding to
dots $0$ and $3$ ($2$ and $4$) in an occupied (empty) state.  It then consists of four steps, see Fig.~\ref{f6},
\begin{equation}\label{protocol1}
|\uparrow\uparrow\rangle_{ab}\xrightarrow{\rm (i)}|\downarrow\uparrow\rangle_{ab}\xrightarrow{\rm (ii)}
|\downarrow\downarrow\rangle_{ab}\xrightarrow{\rm (iii)}|\uparrow\downarrow\rangle_{ab}\xrightarrow{\rm (iv)}
|\uparrow\uparrow\rangle_{ab}.
\end{equation}
After the cycle has been completed, the initial spin configuration is reached again.
In each step, either $\varepsilon_a(t)$ [during steps (i,iii)] or $\varepsilon_b(t)$ [during (ii,iv)] is varied, whereas
the respective other parameter is held constant at large absolute value $|\varepsilon_{b/a}|= D\gg |\hat T_{b/a}|$, i.e.,
$H_{b/a}$ does not influence that part of the protocol. 

The full system then includes the code qubit $(\hat Z,\hat X)$ and the  $(a,b)$ spins, i.e., the two dot pairs.
We start from the initial product state
\begin{equation}\label{initialstate}
|\Psi_0\rangle =(\alpha|0\rangle+\beta|1\rangle)\otimes |\uparrow\uparrow\rangle_{ab}= 
|\psi\rangle\otimes |\uparrow\uparrow\rangle_{ab},
\end{equation}
with arbitrary code qubit state $|\psi\rangle$.  In order to prepare the ancilla state $|A_\theta\rangle$,
one puts $\alpha=\beta=1/\sqrt2$ such that $|\psi\rangle=|+\rangle$, cf.~Eq.~\eqref{ancilla-theta}.
For a study of the subsequent time evolution under the time-dependent steering protocol in Eq.~\eqref{protocol1}, it 
is convenient to introduce the effective parity operators 
\begin{equation}\label{parityspin}
\hat{\cal P}_a = 2S_a^z \hat Z, \quad \hat {\cal P}_b = 2S_b^z \hat Z, \quad \hat {\cal P}_{\rm tot}=4S_a^z S_b^z \hat Z,
\end{equation}
which have eigenvalues ${\cal P}_{a/b/{\rm tot}}=\pm 1$.    We observe that during steps (i,iii) of the above protocol,
  $\hat {\cal P}_a$ is conserved, $[\hat {\cal P}_a,H]=[\hat{\cal P}_a,H_a]=0$, 
since $H_b$ here reduces to an irrelevant constant as explained above. 
 Similarly, $\hat {\cal P}_b$ is conserved during steps (ii,iv).
In addition, the operator $\hat{\cal P}_\mathrm{tot}$ in Eq.~\eqref{parityspin} is conserved 
during the entire protocol, and we can separately consider the parity sectors ${\cal P}_{\rm tot}=\pm 1$.

Let us first discuss the ${\cal P}_\mathrm{tot} = +1$ sector, where one effectively starts  from the 
initial state $|0\rangle \otimes |\uparrow\uparrow\rangle_{ab}$ with amplitude $\alpha$, see Eq.~\eqref{initialstate}. 
(For ${\cal P}_{\rm tot}=-1$, see below.)  Under the protocol (\ref{protocol1}), the effective parities ${\cal P}_{a,b}$ develop as
 \begin{eqnarray} \label{parityprotocol}
 {\cal P}_a &=& +1 \xrightarrow{\rm (i)} +1 \xrightarrow{\rm (ii)}
-1 \xrightarrow{\rm (iii)} -1 \xrightarrow{\rm (iv)}+1,\\  \nonumber
 {\cal P}_b &=& +1 \xrightarrow{\rm (i)} -1 \xrightarrow{\rm (ii)} -1 \xrightarrow{\rm (iii)} +1 \xrightarrow{\rm (iv)}+1.
 \end{eqnarray}
In each step of the protocol, by exploiting conservation of ${\cal P}_{\rm tot}$ and one of the ${\cal P}_{a/b}$ parities,
only two basis states are dynamically relevant.  For instance, during step (i) where ${\cal P}_a=+1$ is conserved, 
those two states are $|0\rangle\otimes \uparrow\uparrow\rangle_{ab}$ and $|1\rangle\otimes|\downarrow\uparrow\rangle_{ab}$.
On the other hand, during step (iii) with conserved parity value ${\cal P}_a=-1$, they are given by
$|0\rangle\otimes|\downarrow\downarrow\rangle_{ab}$ and $|1\rangle\otimes |\uparrow\downarrow\rangle_{ab}$, see Fig.~\ref{f6}.   The physics during step (i,iii) is then 
described by the respective simple $2\times 2$ Hamiltonian 
\begin{equation}\label{HpumpA2}
H_a^{({\cal P}_a = \pm 1)} = \left(\begin{array}{cc} \varepsilon_a & i (t^x_a\mp t^y_a) 
\\ -i (t_a^x \mp t^y_a)^\ast & -\varepsilon_a\end{array}\right).
\end{equation}
The corresponding Hamiltonian for steps (ii,iv) follows by letting $a\to b$ in Eq.~\eqref{HpumpA2}.
Similarly, by taking into account the constraints \eqref{parityprotocol} and the above discussion, 
the state after each of the four steps in Eq.~(\ref{protocol1}) will be given by 
\begin{eqnarray}\nonumber
|0\rangle\otimes |\uparrow\uparrow\rangle_{ab}
\xrightarrow{\rm (i)} && |1\rangle\otimes |\downarrow\uparrow\rangle_{ab}
\xrightarrow{\rm (ii)}  |0\rangle\otimes |\downarrow\downarrow\rangle_{ab}\xrightarrow{\rm (iii)}
 \\  \label{stateprotocol}
\xrightarrow{\rm (iii)}&&  |1\rangle\otimes|\uparrow\downarrow\rangle_{ab}
\xrightarrow{\rm (iv)}|0\rangle \otimes |\uparrow\uparrow\rangle_{ab},
\end{eqnarray}
up to a crucial phase factor discussed next.

We now analyze the steered qubit motion on its Bloch sphere in more detail.
During step (i), spin $a$ is flipped, $|\uparrow\rangle_a\to |\downarrow\rangle_a$, and the qubit state is dragged
from the north to the south pole, $|0\rangle\to |1\rangle$, at azimuthal angle
\begin{equation}\label{deltai} 
\delta^{\rm (i)}_{{\cal P}_a=+} = \arg[-i(t^x_a - t_a^y)^\ast].  
\end{equation}
Since the entire qubit motion proceeds under this angle, the latter can be read off 
from $H_a^{({\cal P}_a=+1)}$ by letting $\varepsilon_a\to 0$,
see Eq.~\eqref{HpumpA2}. In the next step (ii), which is governed by $H_b^{({\cal P}_b=-1)}$,  
spin $b$ is also flipped by a respective gate sweep applied to dot pair ($3,4$). 
Thereby the code qubit is forced to move back to the north pole but under the different azimuthal angle 
\begin{equation}\label{deltaii}
\delta^{\rm (ii)}_{{\cal P}_b=-1} = \arg[-i(t^x_b+ t_b^y)^\ast].
\end{equation}
This difference to Eq.~\eqref{deltai} arises because spin $b$ employs the transfer operator $\hat T_b$ instead of $\hat T_a$.  
We then proceed with step (iii), where spin $a$ returns to the state $|\uparrow\rangle_a$  and thereby drives the qubit 
once more from north to south pole. This happens at the azimuthal angle $\delta^{\rm (iii)}_- = \arg[i(t_a^x + t^y_a)]$. 
In the final step (iv), spin $b$ is also flipped back to $|\uparrow\rangle_b$. With $\delta^{\rm (iv)}_+ = \arg[i(t^x_b-t^y_b)]$, we 
thus have returned to the initial state up to a phase factor.  

The logical qubit state has then been steered twice from north to south pole (and back) on the Bloch sphere. 
Dynamical phases always cancel out under this protocol (see below),  and the phase acquired by the qubit state has
a purely geometric origin. This non-Abelian Berry phase can be 
obtained from the solid angle $\Omega_{{\cal P}_{\rm tot}=+1}$ encircled during a full cycle on the Bloch 
sphere as $\theta = \Omega_{+}/2$ \cite{Pachos1999}. Since we move directly between 
both poles (with different azimuthal angles), one easily obtains 
\begin{equation}\label{solidangle}
\frac{\Omega_{+}}{4\pi} =  \frac{1}{2\pi}\left(\delta^{\rm (i)}_+ - \delta^{\rm (ii)}_- + \delta^{\rm  (iii)}_- -\delta^{\rm (iv)}_+\right).
\end{equation}
The Berry phase can thus be decomposed into two parts depending only on $t^{x,y}_{a}$ or $t^{x,y}_b$ parameters.
With $\theta_a=\delta_+^{\rm (i)}+\delta^{\rm (iii)}_-$ and $\theta_b=-(\delta^{\rm (ii)}_- + \delta^{\rm (iv)}_+)$, 
we find
\begin{eqnarray} \label{thetavar}
\theta &=& \theta_{a}+\theta_b, \\ \nonumber
\theta_{a/b}  &=& \tan^{-1}\left( \frac{2|t^x_{a/b} t^y_{a/b}| } {|t^x_{a/b}|^2-|t^y_{a/b}|^2}\sin\varphi_{a/b} \right),
\end{eqnarray}
where $\varphi_{a/b}$ is the relative phase difference between $t_{a/b}^{x}$ and $t_{a/b}^y$.
As discussed above,  the regime with $|t_b^y/t_b^x|\ll 1$ is naturally expected for the setup 
in Fig.~\ref{f6}, resulting in $\theta_b\to 0$ and hence $\theta \simeq \theta_a$.

To complete the discussion, we now address the other parity sector ${\cal P}_{\rm tot}=-1$, where one starts out
from the initial state $|1\rangle\otimes|\uparrow\uparrow\rangle_{ab}$ with amplitude $\beta$, cf.~Eq.~\eqref{initialstate}.
The corresponding time evolution can be recovered from the above expressions for ${\cal P}_{\rm tot}=+1$
by the replacement ${\cal P}_{a/b}\to -{\cal P}_{a/b}$, implying  
$\delta^{\rm (i,ii,iii,iv)}_+ \leftrightarrow \delta^{\rm (i,ii,iii,iv)}_-$ 
in Eq.~\eqref{solidangle}. Further taking into account that the tunneling processes (i,iii) [and (ii,iv)] are 
related by Hermitian conjugation, i.e.,
\begin{equation}
\delta^{\rm (iii)}_\pm = -\delta^{\rm (i)}_\pm,\quad \delta^{\rm (iv)}_\pm = -\delta^{\rm (ii)}_\pm,
\end{equation} 
the solid angle for ${\cal P}_{\rm tot}=-1$ follows as $\Omega_- = -\Omega_+$, and hence the Berry phase $-\theta$ 
will be picked up. In effect, with $\alpha=\beta=1/\sqrt2$ in Eq.~\eqref{initialstate}, i.e., by preparing the qubit in the $|+\rangle$ state, 
the final state after running the four-step protocol is precisely given by the ancilla state 
$|A_\theta\rangle$ in Eq.~\eqref{ancilla-theta}.  Remarkably, we can thereby realize a static and parametrically tunable phase 
angle $\theta$, see Eq.~\eqref{thetavar}. With $\theta \simeq \theta_a$, a main handle to adjust this angle is given by the backgate
detuning parameter $\Delta n_g$ on the MCB hosting $\gamma_0$, see Fig.~\ref{f6}. Assuming small detuning, $|\Delta n_g| \ll 1$, 
we find $\theta \sim |t^x_a/t^y_a| \sim \Delta n_g E_C^2/\bar t^2$ up to the prefactors in Eq.~\eqref{thetavar}.

By alternative choices of dots for the pump protocol, one may change the parametric dependencies 
of the phase gate angle $\theta$. Using the quantum dot $0\to 0'$ (yellow square) in Fig.~\ref{f6}, the ratio of tunneling 
amplitudes $|t^{x}_{a}/t^{y}_{a}|$ becomes independent of the scale $\bar t/E_c$. Still, now adjusting the backgate parameter $n_g'$ 
of the box hosting $\gamma_0'$ in Fig.~\ref{f6}, one may tune and/or quench the tunneling amplitude $t^{x}_{a}$. 
To leading order, the protocol then yields the phase angle $\theta \sim \Delta n_g'$ while  $\bar t/E_c$ can be changed independently.

Last we address the effect of errors, see also Sec.~\ref{sec4f} below.  
First, energy splittings between the $Z =\pm$ qubit states can result from the static 
stabilizer Hamiltonian \eqref{Sysham} or from the Majorana-dot tunnel couplings entering 
the pumping Hamiltonian, cf.~Eq.~\eqref{HpumpA2}. However, noting that our four-step protocol involves
intermediate parity flips in both subsectors, ${\cal P}_{a/b} \to -{\cal P}_{a/b}$,
we observe that as long as the control parameters $\varepsilon_{a/b}(t)$ run back and forth in a 
time-reversed manner during the respective steps (i,iii) and (ii,iv), 
dynamical phases cancel out automatically. Even under imperfect gate control, see the lower panel of Fig.~\ref{f6}, 
such contributions are expected to echo out.  In particular, by $n>1$ repetitions of a protocol with small phase angle $\theta/n$, 
the Berry phase $\theta$ comes with a more efficient cancellation of dynamical phases.
Second, assuming time-reversed gate sweeps,  the four-step pumping 
sequence (\ref{protocol1}) affords a topological protection in the control parameter 
space $(\varepsilon_a,\varepsilon_b)$, i.e., the pumped phase will be 
independent of the precise time dependence of $\varepsilon_{a/b}(t)$ 
as long as the intermediate steps of the protocol in Eq.~\eqref{protocol1} are reached.
Third, we note that small deviations from adiabaticity are permitted 
if diabatic errors are eliminated by intermediate charge measurements on the dots \cite{Nayak2016}.
We emphasize that the static code parameters yielding $\theta$ in Eq.~\eqref{thetavar} can be adjusted and fine-tuned. 
Since the protocol forms a closed loop in control parameter space, in principle, it can be run indefinitely. 
For instance, after preparing the desired ancilla state $|A_\theta\rangle$,  this state can be post-processed or 
verified by state tomography.  One can then readjust the code parameters in order to optimize the protocol.
Finally, charge dephasing may be a challenge, in particular if the anticrossing energy
$|\hat T_{a/b}|$ is very small.
Extensions of the protocol (cf.~Refs.~\cite{Nayak2016,Karzig2016}) 
should be possible and can be used to tackle some of the challenges outlined here. We leave a detailed investigation to future work.

To conclude, the presented phase gate protocol follows from a gate-steered rotation in the hybrid 
low-energy subspaces of the quantum dots and the code qubit. Such control and rotations are possible since single-electron tunneling events 
generate logical string operators acting on the code space, see Sec.~\ref{sec2c}. As electrons traverse the code, these strings drag along $Z$- 
and $X$-type excitations, which can be created and braided in nontrivial interleaved pumping sequences. In the manipulations applied 
to our code architecture, the coupling to different Pauli operators or strings naturally arises from superpositions of tunneling paths through the code.

\subsection{Errors and challenges}\label{sec4f}

Having discussed a framework of elemental quantum logical operations sufficient to the implementation of gates, 
we are now in position to address likely sources of errors or other physical mechanisms detrimental 
to the functioning of the system. Generally speaking, we expect the topological protection of the elementary Majorana particles 
(on which our approach is based) to manifest itself in long coherence and qubit life times. This may turn out to be an 
important mechanism keeping the need for active error correction at tolerably low levels. Second, a general advantage of the 
stabilizer code approach is that Clifford gates do not require active error correction \cite{Nielsen2000}. 
Measurement errors then merely need to be kept track of, which can be done by classical software \cite{Fowler2012}.  
We thus are confident that the present Majorana-based code can be implemented with a reasonably low operational
overhead. Likely error sources and/or challenges include:

\subsubsection{Dynamical phase errors}

These errors are caused by the effectively random (but static) coupling constants
generated by tunneling processes encircling single or multiple stabilizers. As a consequence,
the time evolution even of stationary states leads to dynamical phases. 
The ensuing errors are correctable if, in each cycle, they affect less than half of the stabilizers defining an 
encoded logical qubit \cite{Fowler2012,Terhal2015}. Generally, it will thus be advantageous to 
reduce or quench off the tunnel couplings contributing to a logical qubit. Increasing the physical to logical qubit ratio 
then exponentially suppresses the remaining detrimental tunneling terms, and therefore provides exponential protection 
against dephasing errors.

\subsubsection{MCB errors}

A second class of dynamical errors has its origin in the less than perfect design of the MCBs. 
The functioning of our approach rests on the approximate degeneracy of the two quantum states
realized by a MCB in the strong charging limit. That qubit degeneracy gets lifted when the Majoranas constituting it
hybridize with each other, e.g., due to the finite length of the  MCB semiconductor quantum wires.  The actual splitting for 
realistic geometries is not yet known and additional experiments on simple Majorana-based qubits will be needed to estimate these error rates. 
In terms of MCB Pauli operators \eqref{PauliMajorana}, such residual overlaps act as effective magnetic fields.  In general, these Pauli
errors are correctable.

A qualitatively different class of errors originates in quasiparticle poisoning. For example, quasiparticles may jump on and off neighboring MCBs.  Such processes are efficiently suppressed by the charging energy, which in practice results in charge stability over very long times (hours).  Nonetheless,
quasiparticle poisoning can have detrimental consequences to the operation of the surface code  
when Cooper pairs on a given MCB are likely to break up --- a process which has been 
experimentally verified to occur at
 time scales beyond 10~ms for single nanowires
 \cite{Higginbotham2015}. If the electrons recombine 
with two different Majorana states, this will create a Pauli error, which is straightforward to detect via an observable change of the
stabilizer states connecting to the MCB. Such Pauli errors can again be corrected.
On the other hand, if only one quasiparticle recombines with a Majorana state, the error is hard to correct since 
the system leaves the computational subspace. Such an event will manifest itself in a flip of the fermion parity 
operator $\hat {\cal P}_l$ on a single MCB, in addition to flips of the adjacent $\hat Z$ and $\hat X$ stabilizers.
This error is not correctable anymore, since we do not read out the operator $\hat {\cal P}_l$ needed to unambiguously detect it
in the present scheme.

Once such an error has occurred, we are left with an out-of-subspace Majorana system plus an isolated quasiparticle.
The latter may remain idle on the MCB for some time. Eventually, the original state of the
system gets restored (vanishing of the error) or recombination with another Majorana state takes place (return to the computational subspace plus a  Pauli error).
If the out-of-subspace time exceeds typical operation times and/or quasiparticle poisoning errors occur on logical qubits, the functioning of the code gets compromised.
Fortunately, errors of this type betray themselves via correlated flips of neighboring stabilizers $\hat Z$ and $\hat X$. 
If they turn out to be important, measurements of the box parity operators $\hat P_l$ should be implemented. However,
this will require a modified approach than presented here.

\subsubsection{Conductance interferometry and readout scheme}

Our conductance readout schemes require reference arms for path interference. 
Quantum coherence will only be preserved if the length of these arms is shorter than the phase coherence length in the 
relevant material. Since the distance between (zero-energy) Majorana states on a given MCB should be larger than 
the coherence length in the topological superconductor \cite{Alicea2012,Leijnse2012}, 
this sets a minimum length for the reference arms as well.  The magnitude of a realistic topological 
gap is not known, but one expects a decay length of Majorana states in the $\mu$m regime \cite{Albrecht2016}. 
However, even if the coherence length of the reference arm is of order $\mu$m, the conductance
measurement has to be done at low voltages to avoid decoherence of interference signals. 
All factors considered, it might be favorable to use a topological wire to extend the coherence of the 
reference arm, similar to the interferometer suggested by Fu \cite{Fu2010}.
(We note that such long links are needed only along the wire direction in Fig.~\ref{f1}.)
Again one will have to rely on experimental progress to determine what the best solution is, 
and therefore we leave this as an open challenge.   
  
Another important issue is to perform stabilizer measurements on a sufficiently fast time scale.
However, the conductance in the readout scheme will be small because the MCBs are
operated in the cotunneling regime, and hence the necessary integration time might be
long. A different option is to read out tunneling times by charge measurements, which
means that one could change the design by replacing the normal leads with dots and
charge sensors \cite{paperB}. This is a known technology with readout times below microseconds
\cite{Petta2005} and offers the advantage that only one dot needs to be coupled to
each Majorana state (instead of a lead and a dot). Also in this case, actual
experimental realizations will determine the best strategy.  An interesting
question, where the answer depends on the actual readout strategy, is how fast a
projective measurement can be. It will also be important to understand better, on a
theoretical level, the time scales for projective measurements when the measured
quantities are tunneling rates rather than the transferred charge.

\subsubsection{2D vs 3D architecture}

When discussing hardware solutions, errors, and their countering, 
one should have in mind that the 3D integration envisioned in Ref.~\cite{Aviad2016}, 
cf.~also Fig.~\ref{f1}, with controllable tunnel junctions connecting the code to another layer
containing  gate electrodes and probe leads, may be difficult to realize. 
This is not a fundamental obstacle, but something that still needs to be developed along 
with the topological wire layouts.
The first demonstration experiments should therefore be designed in a purely 2D setup \cite{paperB}, 
where one can also address charge vs conductance readout and the role of quasiparticle poisoning.

\section{Logical gates for universal quantum computation}\label{sec5}

We now turn to logical gates in the present code architecture, 
which can be implemented by known quantum circuits 
\cite{Fowler2012,VijayFu2015,Nielsen2000,Preskill} once the basal quantum 
operations in Sec.~\ref{sec4} have been realized. 
We consider the set $\{ \hat C, \hat H, \hat S, \hat T\}$ of
elementary gates, containing the CNOT $\hat C$, the
Hadamard $\hat H$, and the phase gates $\hat S$ and $\hat T$. This set 
is sufficient for universal quantum computation \cite{Nielsen2000}. 
Since $\hat T^2= \hat S$, the $\hat S$-gate is formally not necessary for
a minimal set, but access to a simple and efficient realization is highly desired 
in practice. 
 
\subsection{Controlled-NOT gate}\label{sec5a}
 
 We first recall that in the Schr\"odinger picture, the CNOT acts on a two-qubit
 state as $|\psi_A \rangle \otimes |\psi_B\rangle \to \hat C (|\psi_A\rangle
 \otimes|\psi_B\rangle)$, where  the unitary operator
 \begin{equation}\label{eq-CNOT}
\hat C=\frac{(\hat I +\hat Z)_A}{2}\otimes \hat I_B
 +\frac{(\hat I -\hat Z)_A}{2}\otimes \hat X_B
 \end{equation}
 flips the target qubit $B$ iff the control qubit $A$ is in a down-state \cite{Nielsen2000}.
 The Schr\"odinger description of $\hat C$ in Eq.~\eqref{eq-CNOT} is  equivalent 
 to a Heisenberg transformation, $\hat{\cal O}_A\otimes  \hat{\cal O}_B \to \hat C^{-1}(\hat{\cal O}_A\otimes
\hat{\cal O}_B)\hat C$.  This transformation is recovered by the braiding-induced manipulation of 
operators in Eq.~\eqref{braid-op}, see~Sec.~\ref{sec4c}. Since all other transformations of the 
two-qubit operator algebra can be constructed from Eq.~\eqref{braid-op}, 
those four rules are sufficient to fully characterize the CNOT gate \cite{Nielsen2000}.
As shown in Sec.~\ref{sec4c}, cf.~also Fig.~\ref{f4}, the rules in
Eq.~\eqref{braid-op} can be directly implemented by braiding 
a pair of different-type logical qubits. 
We note that the CNOT operation for same-type qubits  
can also be obtained by a variant of our protocol. 
However, since braiding is limited to different-type qubits, one then needs 
additional ancilla qubits serving as target of an intermediate CNOT,
cf.~Refs.~\cite{Fowler2012,VijayFu2015}.
Last, as both qubit move and braid operations allow for straightforward iteration on qubits with larger stabilizer and/or string 
distances ($d > 1$), the corresponding extension of the CNOT follows directly.

\subsection{Hadamard gate}\label{sec5b}

The Hadamard gate applies the operator $\hat H= (\hat X+\hat Z)/\sqrt{2}$
on a single logical qubit state, $|\psi\rangle\to \hat H|\psi\rangle$, 
mapping $Z$-eigenstates $|0\,/1\rangle$ to $X$-eigenstates $|+/-\rangle$ 
(Schr\"odinger) or, equivalently, exchanging $\hat X/\hat Z \to \hat Z/\hat X$ 
(Heisenberg) \cite{Nielsen2000}.
An important advantage of the Majorana surface code comes from the fact that the $\hat H$-gate 
can be realized by moving logical information between adjacent qubits of different ($X/Z$)
type \cite{VijayFu2015}. This process effectively exchanges $\hat X$ and $\hat Z$ operators. 
Fortunately, such moves can be performed in a rather simple manner in 
our code architecture, see Sec.~\ref{sec4b}. The key ingredient is an interferometric 
conductance measurement of a string  operator, as illustrated in Fig.~\ref{f3} 
for the case of minimal string extension (with code distance $d=1$). 
 An alternative scheme employs stabilizer product operators, see also Sec.~\ref{sec4b}.
 
In order to improve error resilience, a $d>1$ generalization of either approach to qubits 
with longer connecting strings and/or larger holes is possible.  In the latter case, sequential 
application of the above manipulations to all constituent elementary qubits implements the 
Hadamard gate.  Similarly, for longer strings, $\approx d$ sequential non-stabilizer 
measurements are needed.  Fault tolerance is then retained because all measured 
operators commute with local pairs of $\hat Z$- and $\hat X$-type stabilizer (or string) 
operators, thus allowing for error checks.

\subsection{S gate} \label{sec5c}

\begin{figure}
\centering
\includegraphics[width=7cm]{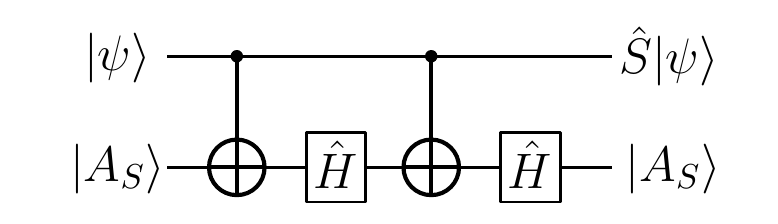}
\caption{Quantum circuit for $\hat S$-gate implementation \cite{Fowler2012}.
We start from $|\Psi\rangle_{\rm initial}=|\psi\rangle\otimes |A_S\rangle$, 
with an arbitrary state $|\psi\rangle$ stored on
the logical qubit and the $Y$-eigenstate 
$|A_S\rangle$, see Eq.~\eqref{ancillaS},
 on the ancilla qubit.  One then applies a CNOT (with the ancilla as target),  
followed by a Hadamard on the ancilla.  Repeating this sequence once more, 
one arrives at $|\Psi\rangle_{\rm final}=
(\hat S|\psi\rangle) \otimes |A_S\rangle$.   \label{f7}  }
\end{figure}

We now turn to the implementation of the $\hat S$-gate, which acts on a single logical qubit 
state $|\psi\rangle$ by creating a $\pi/2$ phase shift between the up- and down-components.
Using the phase gate operator $\hat P(\theta)=e^{i\theta\hat Z}$ in Eq.~\eqref{phasegateop}, 
this is achieved by the operator 
\begin{equation}\label{sgateop}
\hat S=e^{i\pi/4} \hat P(-\pi/4)={\rm diag}(1,i).
\end{equation}
To realize the $\hat S$-gate, we consider the two-qubit protocol in Fig.~\ref{f7} \cite{Fowler2012,Nielsen2000}, 
which employs an ancilla qubit in addition to the logical qubit.  Here the ancilla is prepared in 
the Pauli-$Y$ eigenstate
\begin{equation}\label{ancillaS}
 |A_S\rangle = \hat S|+\rangle = (|0\rangle + i|1\rangle)/\sqrt2,
\end{equation}
or, equivalently, $|\bar A_S \rangle = \hat{S}^\dagger |+\rangle$. Note that 
$\hat Y|A_S\rangle=|A_S\rangle$ and $\hat Y|\bar A_S\rangle=-|\bar A_S\rangle$, where
$|A_S\rangle$ is equivalent to $|A_{\theta=-\pi/4}\rangle$ in Eq.~\eqref{ancilla-theta}.
By applying a sequence of CNOT and Hadamard gates,
the circuit in Fig.~\ref{f7} arrives (in the absence of errors) at the product state $|\Psi\rangle_{\rm final}=(\hat S|\psi\rangle)\otimes |A_S\rangle$, 
containing the desired state $\hat S|\psi\rangle$ on the logical qubit.
At the same time, one recovers the original ancilla state.

The $\hat S$-gate implementation then boils down to a reliable generation of $|A_S\rangle$ ancilla states. 
In principle, these could be obtained by  direct $\hat Y$-measurements on the qubits of the code. 
However, since many ancilla states have to be generated in practice, an attractive alternative 
is offered by the state injection approach in Sec.~\ref{sec4d}. Here one first prepares MCB Pauli-$\hat y$ states on a 
separate auxiliary box. In a second step, this MCB qubit state is transferred (injected) to a selected ancilla qubit in the code.
Given that this preparation scheme relies on projective measurements, the obtained states are 
expected to naturally have high fidelity with $|A_S\rangle$ in Eq.~\eqref{ancillaS} \cite{footnote-sgate},
possibly obviating the need for magic state distillation \cite{Bravyi2005}.  
This route can offer a significant reduction of the large ancilla preparation overhead needed
in other surface code architectures (see~e.g.~Ref.~\cite{Fowler2012} and references therein),
and could thereby pave the way towards an efficient experimental realization of Clifford operations.

\subsection{T gate}\label{sec5d}

With the set $\{ \hat C,\hat H,\hat S\}$ discussed up to now, we can only perform Clifford operations \cite{Nielsen2000}.  
The Gottesman-Knill theorem \cite{Gottesman1998} states that a Clifford-only quantum computer 
offers no quantum speedup since Clifford-based computations can be carried out efficiently on a classical computer \cite{Terhal2015}.
Although such a system can still be highly useful as platform for many QIP tasks, universal quantum computation requires an additional 
non-Clifford gate, for example the $\hat T$-gate \cite{Nielsen2000}. This gate is equivalent to the $\theta=-\pi/8$ phase gate, 
\begin{equation}\label{tgateop}
\hat T=e^{i\pi/8}\hat P(-\pi/8)={\rm diag}(1,e^{i\pi/4}),
\end{equation} 
and is also used to achieve universality in other 
Majorana-based quantum computation approaches, see, e.g., Refs.~\cite{Clarke2016,Karzig2016}.  
Possibly with minor modifications, our proposal, see Sec.~\ref{sec4e} and below, can be useful to (or make use of) those schemes as well.  

\begin{figure}
\centering
\includegraphics[width=7cm]{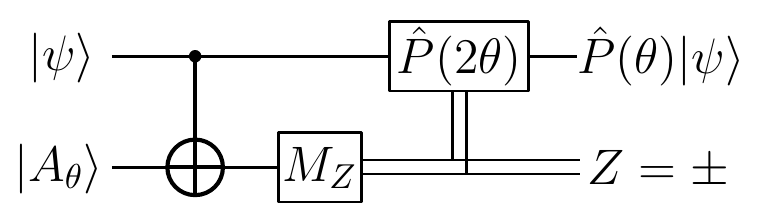}
\caption{ \label{f8} Quantum circuit for implementing a phase gate $\hat P(\theta)$,
see Eq.~\eqref{phasegateop}, with arbitrary angle $\theta$ \cite{Preskill}.
The $\hat T$-gate follows for $\theta=-\pi/8$, see Eq.~\eqref{tgateop}. Starting from the two-qubit state 
$|\Psi\rangle_{\rm initial}=|\psi\rangle\otimes|A_\theta\rangle$,
with arbitrary state $|\psi\rangle$ on the logical qubit and 
the ancilla state $|A_\theta\rangle$ in Eq.~(\ref{ancilla-theta}),
a CNOT is followed by a Pauli-$Z$ measurement ($M_Z$) 
on the ancilla (outcome $Z=\pm 1$). Double lines indicate 
classical information processing, where $\hat P(2\theta)$ is applied to 
the logical qubit iff $Z = -1$. The final logical qubit state is $\hat P(\theta)|\psi\rangle$.
}
\end{figure}

Our phase gate implementation uses the quantum circuit in Fig.~\ref{f8}, which entangles a logical qubit 
containing an arbitrary state $|\psi\rangle$ with an ancilla qubit state $|A_\theta\rangle$.
The latter follows by applying $\hat P(\theta)$ to the $|+\rangle$-eigenstate of the ancilla, see Eq.~\eqref{ancilla-theta}.
After a CNOT, one measures the ancilla Pauli-$Z$ component (outcome $Z=\pm 1$).
For $Z=+1$, one directly obtains the desired state $\hat P(\theta)|\psi\rangle$ on the logical qubit.
However, for $Z=-1$, one obtains $\hat P(-\theta)|\psi\rangle$ and then has to apply the operator $\hat P(2\theta)$ 
to correct the result. 
Since for $\theta=-\pi/8$, the operator $\hat P(2\theta)$ represents the relatively simple Clifford operation $\hat S$ in Eq.~\eqref{sgateop}, 
when using the circuit in Fig.~\ref{f8}, the $\hat T$-gate will be much easier to realize than arbitrary-$\theta$ phase gates.  
The above protocol also can be used to implement the $\hat S$-gate ($\theta=-\pi/4$),
where $\hat P(2\theta)$ is equivalent to the Pauli operator $\hat Z$.
However, since one here loses the ancilla state, the $\hat S$-gate implementation in Sec.~\ref{sec5c} is more 
efficient if ancilla preparation constitutes the bottleneck. 
  
In order to implement the $\hat T$-gate in our code architecture, one may 
 first generate the ancilla states $|A_{\theta=-\pi/8}\rangle$  by the multi-step electron pumping protocol
 in Sec.~\ref{sec4e}.  These states are then processed 
 by the circuit in Fig.~\ref{f8}, which yields the desired logical qubit gate, $\hat T|\psi\rangle$.
For elementary QIP tasks and first experimental checks, we expect that 
our pumping protocol can directly yield ancilla states of sufficiently high fidelity. 
In such basic cases, phase gate operations may directly be applied to logical qubit states without the 
need for additional magic state distillation steps \cite{Bravyi2005}, thus resulting   
in a significantly reduced overhead in both hardware and operation.
The above route towards tunable phase gates also works for setups composed of a 
few elementary MCB qubits \cite{paperB}.

\section{Conclusions}\label{sec6}

We have discussed a Majorana surface code platform based on topological semiconductor qubits, 
where the nodes of the network are Majorana-Cooper boxes hosting four Majorana bound states. 
The fermion parity of each MCB is fixed by the charging energy, leading to a two-level 
(spin-$1/2$) system protected by the nonlocal properties of Majorana states. 
These nonlocal properties give a natural anisotropic coupling \cite{Terhal2012} between 
neighboring MCBs, which is ideal for a surface code implementation
where stabilizers, consisting of products of $\hat{x}$ or $\hat{z}$ Pauli operators surrounding a 
plaquette, are being measured. Building on this platform, we here took the crucial next step to introduce concrete and arguably realistic ways of operating this surface code architecture. 
It is worth noting that the Majorana surface code avoids the use of extra 
measurement qubits as in bosonic codes \cite{Fowler2012,Martinis2015,Corcoles2015,Diamond2015}, 
where our readout protocols are based on interfering electron paths and employ purely electrical measurements as opposed to readout via coupling to microwave photons (see, e.g., Refs.~\cite{VijayFu2015,VijayFu2016}).  

The surface code architecture  presented here offers a fault-tolerant set of gates, including CNOT and Hadamard. To complete the group of Clifford gates, we have shown how the $\hat S$-gate can be implemented via state injection of an ancilla state 
into the code, which for advanced QIP tasks may be augmented by a magic state distillation protocol. 
In order to make the surface code computation universal, one needs a single non-Clifford gate, for example a phase gate with arbitrary phase
angle. We here take advantage of the unique characteristics of the Majorana surface code hardware to propose 
qualitatively new implementations of gate operations in terms of a multi-step single-electron pumping protocol.
This protocol has the interesting property that it always gives a phase gate, where
the phase angle of the gate is protected in the sense that it depends on fixed and tunable tunnel coupling amplitudes but not on the detailed timing of the protocol, cf.~Sec.~\ref{sec4e}.
In this respect, it is similar to other schemes for adiabatic non-Clifford gates \cite{Flensberg2011,Clarke2016}.

In view of currently unknown experimental perspectives, the realization of hybrid platforms comprising different 
types of Majorana based architectures 
\cite{Sau2010,Alicea2011,Clarke2011,Flensberg2011,Leijnse2011,Bonderson2011,Hyart2013,Pedrocchi2015,Clarke2016,Loss2016}
may offer practical advantages.  With the state injection protocol in Sec.~\ref{sec4d}, we have proposed a route to couple other 
systems to the Majorana surface code or to related setups composed of MCB-based qubits.
To investigate the feasibility of our surface code proposal, first experimental tests should focus on select building blocks of the full setup.  
This will allow one to characterize important system parameters, e.g., life times of MCB qubits, readout times, and so on. 
Among the first steps could be to demonstrate error correction in
devices containing a few qubits, similar to what has been done for bosonic codes with superconducting qubits \cite{Barends2014,Jeffrey2014,Corcoles2015,Martinis2015}. Besides considering specific error correction schemes, such tests would also be necessary to investigate the feasibility of using MCBs as fundamental qubits
in simpler system containing a few boxes. In a forthcoming publication \cite{paperB}, we will address the prospects for QIP in few-MCB systems 
in a design reasonably close to currently realized Majorana devices \cite{Higginbotham2015,Albrecht2016}.

To conclude, we hope that our proposal will stimulate further experimental and theoretical work towards 
quantum computation based on Majorana systems. 
In particular, we believe that the idea of using parity-fixed Majorana boxes with four Majorana states as elementary qubits opens very promising perspectives in this context.

\acknowledgments
We thank S. Albrecht, P. Bonderson,  M. Freedman, L. Fu, D. Gross, F. Hassler, 
H. Kampermann, C. Marcus, and A. Rasmussen for inspiring discussions. 
We acknowledge funding by the Deutsche Forschungsgemeinschaft (Bonn) 
within CRC network TR 183  (A.A., K.F., R.E.), by the Danish National Research 
Foundation and Microsoft Research (K.F.), by the Israel Science Foundation 
Grant No.~1243/13 (E.S.), and by the Marie Curie CIG Grant No.~618188 (E.S.).


\begin{thebibliography}{99}
\bibitem{footnote-sgate}
Errors may occur due to poor fault tolerance of the ancilla code qubit, 
which is small and thus has low code distance $d$, cf.~Fig.~\ref{f5}. 
After the injection, however, one could switch to bigger $d$ and thereby 
avoid further errors.
\end{thebibliography}
\end{document}